\documentclass[aps, prd, onecolumn, tightenlines, notitlepage, superscriptaddress, nofootinbib, preprintnumbers, floatfix,showkeys,11pt,altaffilletter]{revtex4-2}

\usepackage{upgreek}

\usepackage[official]{eurosym}
\usepackage[normalem]{ulem}
\usepackage{amstext}
\usepackage{graphicx}
\graphicspath{{figures/}}
\usepackage{url}
\usepackage{color}
\usepackage{ulem}
\usepackage[version=4]{mhchem}
\usepackage[utf8]{inputenc}
\usepackage{fontawesome}
\usepackage{yfonts}

\pdfoutput=1
\usepackage{textcomp}
\usepackage{comment}
\usepackage{yfonts}
\usepackage{epsfig,amsfonts,mathrsfs,graphicx,color,slashed,multirow}
\usepackage{latexsym,graphicx,slashed,color,enumerate,url,cancel,gensymb}
\usepackage{textcomp}

\usepackage[x11names]{xcolor}
\usepackage[colorlinks,pdfstartview=FitV,breaklinks=true]{hyperref}
\usepackage{booktabs}
\usepackage{adjustbox}
\usepackage{textgreek} 
\usepackage{caption, subcaption} 
\usepackage{booktabs} 
\usepackage{float}

\usepackage{lmodern}
\usepackage{ae,aecompl}
\usepackage{appendix}
\usepackage{orcidlink}
\usepackage{nccmath} 
\usepackage{textcomp}
\allowdisplaybreaks 


\def\cevns{CE\textnu NS}
\def\eves{E\textnu ES}
\def\d{\mathrm{d}}
\newcommand{\qtransfer}{\left|\mathbf{q}\right|}

\definecolor{vdrgreen}{rgb}{0.0, 0.6, 0.0}

\definecolor{byzantium}{rgb}{0.44, 0.16, 0.39}

\makeatletter
    \newcommand{\colorboxed}[3][white]{\fcolorbox{#2}{#1}{\m@th$\displaystyle#3$}}
\makeatother

\AtBeginDocument{\hypersetup{citecolor=byzantium,linkcolor=byzantium,urlcolor=byzantium}}
\usepackage{appendix}

\begin{document}

\title{{\LARGE Bounds on new neutrino interactions from\\the first \cevns~data at direct detection experiments}}

\author{Valentina De Romeri~\orcidlink{0000-0003-3585-7437}}
\email{deromeri@ific.uv.es}
\affiliation{Instituto de F\'{i}sica Corpuscular (CSIC-Universitat de Val\`{e}ncia), Parc Cient\'ific UV C/ Catedr\'atico Jos\'e Beltr\'an, 2 E-46980 Paterna (Valencia) - Spain}

\author{Dimitrios K. Papoulias~\orcidlink{0000-0003-0453-8492}}\email{dipapou@ific.uv.es}
\affiliation{Instituto de F\'{i}sica Corpuscular (CSIC-Universitat de Val\`{e}ncia), Parc Cient\'ific UV C/ Catedr\'atico Jos\'e Beltr\'an, 2 E-46980 Paterna (Valencia) - Spain}

\author{Christoph A. Ternes~\orcidlink{0000-0002-7190-1581}}
\email{christoph.ternes@lngs.infn.it}
\affiliation{Istituto Nazionale di Fisica Nucleare (INFN), Laboratori Nazionali del Gran Sasso, 67100 Assergi, L’Aquila (AQ), Italy
}

\keywords{new interactions, dark matter detectors, solar neutrinos, \cevns}

\begin{abstract}
 Recently, two dark matter direct detection experiments have announced the first indications of nuclear recoils from solar $^8$B neutrinos via coherent elastic neutrino-nucleus scattering (CE$\nu$NS) with xenon nuclei. These results constitute a turning point, not only for dark matter searches that are now entering the \textit{neutrino fog}, but they also bring out new opportunities to exploit dark matter facilities as neutrino detectors. We investigate the implications of recent data from the PandaX-4T and XENONnT experiments on both Standard Model physics and new neutrino interactions. We first extract information on the weak mixing angle at low momentum transfer. Then, following a phenomenological approach, we consider Lorentz-invariant interactions (scalar, vector, axial-vector, and tensor) between neutrinos, quarks and charged leptons. Furthermore, we study the $U(1)_\mathrm{B-L}$ scenario as a concrete example of a new anomaly-free vector interaction. We find that despite the low statistics of these first experimental results, the inferred bounds are in some cases already competitive.  For the scope of this work we also compute new bounds on some of the interactions using CE$\nu$NS data from COHERENT and electron recoil data from XENONnT, LUX-ZEPLIN,  PandaX-4T, and  TEXONO. It seems clear that while direct detection experiments continue to take data, more precise measurements will be available, thus allowing to test new neutrino interactions at the same level or even improving over dedicated neutrino facilities.
 
\end{abstract}
\maketitle


\section{Introduction}

Coherent elastic neutrino-nucleus scattering (\cevns) is a neutral-current process in which a low-energy neutrino scatters off an entire nucleus~\cite{Abdullah:2022zue}. Its first theoretical prediction and the principles for its detection date back to the 1970s and 1980s~\cite{Freedman:1973yd,Drukier:1984vhf}. One main feature of \cevns~is that its Standard Model (SM) cross section is large compared to other neutrino scattering cross sections, as it is coherently enhanced being proportional to the number of nucleons squared. On the other hand, its experimental detection faces technological difficulties, as it requires the observation of nuclear recoils of very low energy. For this reason, this process evaded detection during many decades until its first observation by the COHERENT Collaboration~\cite{COHERENT:2017ipa}, using an intense spallation source producing neutrinos from pions decaying at rest. Further observations and evidence using different targets~\cite{COHERENT:2020iec,COHERENT:2021xmm,Adamski:2024yqt} or a reactor source~\cite{Colaresi:2022obx} have provided valuable information on the \cevns~cross section and its dependence on $N^2$. 

The possibility that \cevns~detectors could be used as dark matter (DM)~\cite{Bertone:2016nfn} detectors was pointed out by Goodman and Witten~\cite{Goodman:1984dc}, anticipating the same experimental challenges. An intense experimental program~\cite{Schumann:2019eaa,Billard:2021uyg} has followed this early suggestion, culminating in current ton-scale DM direct detection experiments. The latest generation of low-threshold dual-phase liquid xenon (LXe) detectors, including the XENONnT~\cite{XENON:2023cxc}, LUX-ZEPLIN (LZ)~\cite{LZ:2022lsv}, and PandaX-4T~\cite{PandaX-4T:2021bab} experiments, has reached impressive sensitivities, although without indicating any conclusive evidence of DM detection. As a by-product, the increase in target size has allowed these facilities to be sensitive to fluxes of astrophysical neutrinos. 
As anticipated, the improvement in the energy threshold at these experiments has now led to detectable rates of $^8$B solar neutrinos inducing \cevns ~\cite{Monroe:2007xp,Vergados:2008jp,Strigari:2009bq}. Neutrino backgrounds from natural~\cite{Billard:2013qya} and artificial~\cite{AristizabalSierra:2024smb}  sources do pose a challenge for DM searches in the form of a \textit{neutrino fog}~\cite{OHare:2021utq}, but at the same time they provide new opportunities to probe the neutrino sector~\cite{Harnik:2012ni,AtzoriCorona:2022jeb,deGouvea:2021ymm,Giunti:2023yha,Cerdeno:2016sfi,Dutta:2017nht,Gelmini:2018gqa, Essig:2018tss,Boehm:2020ltd,AristizabalSierra:2020edu, AristizabalSierra:2020zod,Amaral:2020tga,Dutta:2020che,Amaral:2021rzw,DeRomeri:2024dbv,Aalbers:2022dzr,Alonso-Gonzalez:2023tgm,Amaral:2023tbs,Majumdar:2024dms}.

A steady increase in sensitivity has allowed the XENONnT~\cite{XENON:2024ijk} and PandaX-4T~\cite{PandaX:2024muv} Collaborations to recently report their first indications of nuclear recoils from solar $^8$B neutrinos via \cevns. With the present exposures reached in these experiments, the background-only hypothesis is disfavored with a statistical significance of $2.73 \sigma$ in the case of XENONnT and $2.64 \sigma$ for PandaX-4T.
These results update previous searches by the same Collaborations~\cite{XENON:2020gfr,PandaX:2022aac} and constitute the first indication of nuclear recoils from solar neutrinos and the first \cevns~observation on a xenon target.
Assuming that no new physics is present, these results also provide a measurement of the $^8$B solar neutrino flux component which is in agreement with theoretical predictions~\cite{Vinyoles:2016djt} and with results from dedicated neutrino experiments~\cite{SNO:2011hxd,KamLAND:2011fld,Borexino:2017uhp,Super-Kamiokande:2016yck}.
These successful experimental results have immediately motivated new phenomenological studies, aiming at probing new physics in the form of non-standard neutrino interactions (NSI)~\cite{AristizabalSierra:2024nwf,Li:2024iij}, light mediators~\cite{Xia:2024ytb} and also the determination of the weak mixing angle at low momentum transfer~\cite{Maity:2024aji}.

In this paper, we study some implications of the first \cevns~indications at DM experiments both for SM and new physics. Following Ref.~\cite{Maity:2024aji}, we first revisit the determination of the weak mixing angle at the $\mathcal{O}(10)$ MeV scale, providing also a result in terms of a combined (XENONnT and PandaX-4T) analysis. Next, we confront new interactions between neutrinos, quarks and charged leptons with the recent XENONnT and PandaX-4T data. We focus on Lorentz-invariant interactions (scalar, vector, axial-vector and tensor), parameterized in a model-independent way in the form of neutrino generalized interactions (NGIs)~\cite{Lee:1956qn,Rodejohann:2017vup,Lee:1956qn,Lindner:2016wff,AristizabalSierra:2018eqm,Chen:2021uuw}. We consider both regimes of effective operators and light mediators, since direct detection experiments have low-energy thresholds and are hence sensitive to interactions involving light mediators~\cite{Cerdeno:2016sfi,Bertuzzo:2017tuf, Farzan:2018gtr, Denton:2022nol}. In addition, for the sake of example, we consider a motivated, anomaly-free $U(1)$ extension of the SM with a B-L symmetry (B being the baryon
number and L the total lepton number).

The remainder of this paper is organized as follows. 
In Sec.~\ref{sec:CEvNS} we introduce the relevant \cevns~cross sections, both in the SM and in the presence of NGIs. We discuss in Sec.~\ref{sec:exp} the simulation details as well as the procedure followed for the statistical analysis of XENONnT and PandaX-4T data. We present in Sec.~\ref{sec:results} our results in terms of a determination of the weak mixing angle at low energy and exclusion limits on the NGI parameter space. Finally, we draw our conclusions in Sec.~\ref{sec:concl}.

\section{Coherent elastic neutrino-nucleus scattering cross sections}
\label{sec:CEvNS}

In this section, we provide the relevant \cevns~cross sections, in the SM and in the presence of NGIs, required for the calculation of the corresponding event rates.

\subsection{\cevns~cross section in the Standard Model}
\label{subsec:CEvNS-SM}

In the SM, the \cevns~differential cross section with respect to the nuclear recoil energy $T_\mathcal{N}$, neglecting $T_\mathcal{N}/m_\mathcal{N}$ and higher order $\mathcal{O}(T_\mathcal{N}^2)$ terms,  reads~\cite{Freedman:1973yd,Barranco:2005yy}
\begin{equation}
\label{eq:xsec_CEvNS_SM}
\left. \frac{d\sigma_{\nu \mathcal{N}}}{dT_\mathcal{N}}\right|^\mathrm{SM}=\frac{G_F^2 m_\mathcal{N}}{\pi}\left({Q_V^\mathrm{SM}}\right)^2 F_{W}^2(\qtransfer^2)\left(1-\frac{m_\mathcal{N} T_\mathcal{N}}{2E_\nu^2} - \frac{T_\mathcal{N}}{E_\nu} \right) \, ,
\end{equation}
with $G_F$ being the Fermi constant, $E_\nu$ the incoming neutrino energy, while $m_\mathcal{N}$ is the nuclear mass and $Q_V^\text{SM}$ denotes the SM weak charge which is given by 

\begin{equation}
\label{eq:CEvNS_SM_Qw}
    Q_V^\text{SM} = g_V^p Z + g_V^n N \, ,
\end{equation}
where $Z~(N)$ is the proton (neutron) number, and the proton and neutron couplings (at tree level)\footnote{At higher orders these factors become flavor-dependent; the correction to $g_V^n$ is very small, while the correction to $g_V^p$ is quite significant~\cite{Cadeddu:2020lky}, although the proton coupling remains very small in comparison to its neutron counter part.} are given by $g_V^p = (1- 4 \sin^2 \theta_W)/2$ and $ g_V^n = -1/2$, respectively. The weak charge encodes the dependence on the weak mixing angle $\theta_W$ through the proton contribution. From RGE extrapolation, its value in the low-energy regime is expected to be
$\sin^2 \theta_W=0.23857(5)$~\cite{ParticleDataGroup:2024cfk}. Nuclear physics corrections are included in the form factor $F_{W}^2(\qtransfer^2)$, to account for
the finite nuclear spatial distribution.  Given the small momentum transfer involved in the \cevns~of $^8$B solar neutrinos, the dependence on the form factor is small. At the scope of the numerical calculations, we rely on the Klein-Nystrand parametrization~\cite{Klein:1999qj} 

\begin{equation}
  \label{eq:KNFF}
   F_{W}(\qtransfer^2)=3\frac{j_1(\qtransfer R_A)}{\qtransfer R_A} \left(\frac{1}{1+\qtransfer^2a_k^2} \right)\ ,
\end{equation}
where $j_1(x)=\sin(x)/x^2-\cos(x)/x$ is the spherical Bessel function of order one, $a_k = 0.7$~fm and $R_A = 1.23 \, A^{1/3}$ indicates the root mean square (RMS) radius (in [fm]), $A$ being the atomic mass number. The expected magnitude of the momentum transfer is $\qtransfer = \frac{\sqrt{2  m_\mathcal{N} T_\mathcal{N}}}{197.327}~\mathrm{fm^{-1}} \sim \mathcal{O}(10)$ MeV. 
 
\subsection{\cevns~cross section with neutrino generalized interactions}
\label{subsec:CEvNS-NGI}

One of our goals in the present work is to explore the implications of the recent XENONnT and PandaX-4T data on new neutrino interactions. For simplicity, we adopt a phenomenological approach and consider all possible Lorentz-invariant low-energy neutral-current interactions parameterized through the following effective Lagrangian 

\begin{equation}
\label{eq:NGIlagr}
\mathscr{L}^\mathrm{NGI}_\mathrm{NC}  \supset \frac{G_F}{\sqrt{2}}
   \sum_{\substack{a=(S,P,V,A,T),\\\ell = e, \mu, \tau}} C_a  \left(\bar{\nu}_\ell \Gamma^a P_L \nu_\ell \right) \left(\bar N \Gamma_a N\right) \, ,
\end{equation}
where $\Gamma^a=\{\mathbb{I},i\gamma^5, \gamma^\mu,\gamma^\mu\gamma^5,\sigma^{\mu\nu}\}$ (with $ \sigma^{\mu\nu}=i[\gamma^\mu,\gamma^\nu]/2$), $P_L \equiv (1 - \gamma^5) / 2$ is the left-handed projector and $N$ denotes the nucleus.  The $C_a$ are dimensionless coefficients which denote the corresponding neutrino-nucleus couplings for all interactions: scalar ($S$), pseudoscalar ($P$), vector ($V$), axial-vector ($A$) and tensor ($T$).
Notice that we consider only flavor-independent interactions and hence assume that the coupling $C_a$ is the same for each neutrino flavor. Therefore, in the present analysis we do not need to include neutrino oscillations, which ---in addition to the interactions--- would also be modified by flavor-dependent interactions due to matter effects in the Sun, relevant for the energies typical of $^8$B neutrinos.

Because of the typical momentum transfer involved in the \cevns~of solar neutrinos, we aim to extend our phenomenological study to the case of interactions involving light mediators (i.e., with a mass $\mathcal{O} (10)$ MeV). At this scope, we modify the effective-interaction couplings 
by introducing an explicit dependence on the mediator mass $m_a$ arising from the propagator. Consequently, the differential \cevns~cross sections for the NGI interactions read~\cite{Candela:2024ljb}

\begin{align}
    \left.\dfrac{\d \sigma_{\nu \mathcal{N}}}{\d T_\mathcal{N}}\right|^{S} (E_\nu, T_\mathcal{N}) =& \, \dfrac{m_\mathcal{N} C_S^4}{4\pi (m_S^2 + 2m_\mathcal{N} T_\mathcal{N})^2} F_W^2(\qtransfer^2) \dfrac{m_\mathcal{N} T_\mathcal{N}}{E_\nu^2} , \label{eq:cross-section-scalar-CEvNS} \\[4pt]
    \left.\dfrac{\d \sigma_{\nu \mathcal{N}}}{\d T_\mathcal{N}}\right|^{V} (E_\nu, T_\mathcal{N}) =& \, \left[1+ \kappa \frac{C_V}{\sqrt{2}G_F Q_V^\mathrm{SM}\left(m_{V}^2+2 m_\mathcal{N} T_\mathcal{N}\right)}\right]^2 \left.\frac{d\sigma_{\nu_\ell \mathcal{N}}}{dT_\mathcal{N}}\right|^\mathrm{SM}
     \label{eq:cross-section-vector-CEvNS} \\[4pt]
    \left.\dfrac{\d \sigma_{\nu \mathcal{N}}}{\d T_\mathcal{N}}\right|^{A} (E_\nu, T_\mathcal{N}) =& \,  \dfrac{2 m_\mathcal{N}}{2J + 1} \dfrac{g_A^4}{(m_A^2 + 2 m_\mathcal{N}  T_\mathcal{N})^2} 
        \left(
            2 + \dfrac{m_\mathcal{N}  T_\mathcal{N}}{E_\nu^2} - \dfrac{2 T_\mathcal{N}}{E_\nu} \right)   
     \tilde{S}^{\mathcal{T}}(\qtransfer^2)  \, , \label{eq:cross-section-axialvector-CEvNS} \\[4pt]
   \left.\dfrac{\d \sigma_{\nu \mathcal{N}}}{\d T_\mathcal{N}}\right|^{T} (E_\nu, T_\mathcal{N}) =& \, \dfrac{m_\mathcal{N} }{2J + 1} \dfrac{g_T^4}{(m_T^2 + 2 m_\mathcal{N} T_\mathcal{N})^2}  \nonumber \\[4pt]
    &\hspace{2em}
    \times \left[ \left(2 - \dfrac{m_\mathcal{N} T_\mathcal{N}}{E_\nu^2} - \dfrac{2 T_\mathcal{N}}{E_\nu} \right) \tilde{S}^\mathcal{T}(\qtransfer^2)
 + \left(1 - \dfrac{T_\mathcal{N}}{E_\nu} \right)  \tilde{S}^\mathcal{L}(\qtransfer^2)  \right]\,. \label{eq:cross-section-tensor-CEvNS}
\end{align}
Note that the cross section for the vector interactions depends on the specific SM extension under consideration: $\kappa= 1$ in the universal scenario, while $\kappa=-1/3$ in the $\mathrm{B-L}$ model~\cite{Langacker:2008yv,Okada:2018ktp}. The axial-vector and tensor cross section are written directly in terms of the fundamental coupling $g_a$, while 
the couplings $C_a$ in the scalar and vector cross sections can be related to $g_a$ at the quark level following the procedure of DM detection \cite{Cirelli:2013ufw,DelNobile:2021wmp} and are given by

\begin{align}
\label{eq:couplingsCa}
    C_S^2 &\equiv g_S^2 \left( Z \sum_{q = u, d} \dfrac{m_p}{m_q} f_{T_q}^{(p)} + N \sum_{q = u,d} \dfrac{m_n}{m_q} f_{T_q}^{(n)} \right), \\[4pt]
    C_V^2 &\equiv 3A g_V^2 \label{eq:couplingsCV} \, .
\end{align}
Throughout this work, we assume that the new mediator $a$ couples with equal strength to neutrinos, quarks and charged leptons. Under this assumption, $g_a$ is defined as $g_a=\sqrt{g_{\nu a}g_{qa}}=\sqrt{g_{\nu a}g_{\ell a}}$, where $g_{\nu a}$, $g_{qa}$ and $g_{\ell a}$ are the couplings between the mediator and neutrinos, quarks and leptons, respectively. In the previous expressions, $m_p$ and $m_n$ denote the proton and neutron masses, respectively, and $m_q$ are the quark $q$ masses, while $f_{T_q}^{(p)}$ and $f_{T_q}^{(n)}$ represent the quark mass contributions to the nucleon (proton and neutron) mass. Note that the expressions for the axial-vector and tensor mediated cross sections are spin dependent. For the latter two we have explicitly extracted their dependence on the total angular momentum, $J$, of the
nucleus in the ground state. In the case of xenon nuclei, only the $^{129}$Xe and $^{131}$Xe isotopes have spin different from zero ($J^\mathrm{^{129}Xe} = 1/2$ and $J^\mathrm{^{131}Xe} = 3/2$), and therefore induce non-zero axial and tensor contributions. The respective abundances are $26.4\%$ and $21.2\%$. The spin structure functions $\tilde{S}^\kappa(\qtransfer^2)$, where $\kappa = \mathcal{L},\, \mathcal{T}$, account for longitudinal and transverse multipoles calculated using the Shell Model, and have been obtained following~\cite{Hoferichter:2020osn} as explained in Appendix B of~\cite{Candela:2024ljb}\footnote{Note that for the axial-vector interaction the longitudinal contribution is negligible in the \cevns~case~\cite{Hoferichter:2020osn}, contrary to the upscattering scenario studied in~\cite{Candela:2024ljb}.}.
The pseudoscalar interaction is not considered in the following as it turns out to be negligible~\cite{Hoferichter:2020osn} for two reasons: first, its cross section is nuclear-spin suppressed and secondly, it is proportional to $\frac{T_\mathcal{N}^2}{2E_\nu^2}$, and hence also  kinematically suppressed.

\section{Statistical analysis}
\label{sec:exp}

We now proceed to discuss the implementation of the statistical analysis. We analyze the experimental data presented in Refs.~\cite{XENON:2024ijk,PandaX:2024muv}. Both experiments utilize a dual-phase time-projection-chamber (TPC), and produce both scintillation photons (S1 signal) and ionization electrons (S2 signal). In the case of PandaX-4T two data sets were analyzed by the Collaboration, one corresponding to a paired S1 and S2 signal and one using only S2 (referred to as US2). In this paper, we use only the second data set (US2), since not enough information is provided by the experimental Collaboration for an accurate reproduction of the paired data. 

The differential event rate is obtained by a convolution of the neutrino flux with the \cevns~cross section

\begin{equation}
    \dfrac{dR^\mathrm{X,P}}{dT_\mathcal{N}} = \mathcal{A}^\mathrm{X,P}(T_\mathcal{N})\int dE_\nu \dfrac{d\phi}{dE_\nu}\dfrac{\d \sigma_{\nu \mathcal{N}}}{\d T_\mathcal{N}}\,,
\end{equation}
where $\dfrac{\d \sigma_{\nu \mathcal{N}}}{\d T_\mathcal{N}}$ refers to any of the expressions in Eq.~\eqref{eq:xsec_CEvNS_SM} or Eqs.~\eqref{eq:cross-section-scalar-CEvNS}--\eqref{eq:cross-section-tensor-CEvNS}, and $\mathcal{A}^\mathrm{X,P}(T_\mathcal{N})$ is the experiment-dependent efficiency (X stands for XENONnT and P for PandaX-4T) which has been extracted from Refs.~\cite{XENON:2024ijk,XENON:2024hup} and Ref.~\cite{PandaX:2024muv} for XENONnT and PandaX-4T, respectively. The flux of $^8$B solar neutrinos, $\dfrac{d\phi}{dE_\nu}$, is taken from Refs.~\cite{bahcall_web,Bahcall:1996qv} with the normalization defined in  Ref.~\cite{Vinyoles:2016djt} ($5.46 \times 10^6~\mathrm{cm^{-2}~s^{-1}}$).  The data in Refs.~\cite{XENON:2024ijk,PandaX:2024muv} is presented in bins of S2 (number of electrons, $N_{e^-}$) in the case of XENONnT (PandaX-4T). Therefore, the events per bin are given by

\begin{equation}
    R_i^\mathrm{X,P} = c_i\mathcal{E}^{\mathrm{X,P}}\int_i \dfrac{dR^\mathrm{X,P}}{dn^\mathrm{X,P}}dn^\mathrm{X,P}\,,
\label{eq:ev_rate}    
\end{equation}
where the integral is performed over the size of bin $i$, while $n^\text{X}=\text{S2}$ and $n^\text{P}=N_{e^-}$ for the case of XENONnT and PandaX-4T, respectively. Following the Collaborations,  for XENONnT we consider 3 bins in the range [120, 500] photoelectrons (PE), while for PandaX-4T we consider 8 bins in the range [4, 8] $N_{e^-}$~\footnote{These values correspond to a nuclear recoil energy range of [0.97, 5.10]~keV in the case of XENONnT and [0.66, 1.19]~keV for PandaX-4T.}. Note that our simulations do not account for resolution effects since no information about them is provided in the experimental papers. However, even without smearing, we are able to reproduce reasonably well the predicted event rates. 
Even under the same conditions (e.g. same assumptions on flux normalizations), we need to include the correction factors $c_i$ in Eq.~\eqref{eq:ev_rate} in order to match our predictions with the best fit spectra presented in the experimental papers, see the first panel of Fig.~2 in Ref.~\cite{XENON:2024ijk} for XENONnT and the upper panel of Fig.~5 in Ref.~\cite{PandaX:2024muv} for PandaX-4T. 
These factors can be seen as effective efficiencies, included because we are performing a simplified analysis compared to what is done by the Collaborations. Indeed, we only use information on S2, while the experimental analyses rely on many more variables that are fitted simultaneously in a correlated way. 
The inclusion of these factors has nonetheless little effect on the NGI analyses, while in the case of the SM analyses it helped to  better reproduce the results from the experimental Collaborations. 
Going back to  Eq.~\eqref{eq:ev_rate}, $\mathcal{E}^{\mathrm{X,P}}$ is the exposure at each experiment, i.e., $3.51$ t$\times$y for XENONnT and $1.04$ t$\times$y  for PandaX-4T (US2), while the differential event rates are expressed through a change of variables according to 
\begin{equation}
    \dfrac{dR^\mathrm{X,P}}{dn^\mathrm{X,P}} = \dfrac{dR^\mathrm{X,P}}{dT_\mathcal{N}} \dfrac{dT_\mathcal{N}}{dn^\mathrm{X,P}}\, .
    \label{eq:T2n_varchange}
\end{equation}
In the case of XENONnT, the translation between nuclear recoil energy and the S2 signal is carried out through
\begin{equation}
    n^\mathrm{X} = \text{S2} = T_\mathcal{N}Q_y^\mathrm{X}(T_\mathcal{N}) g_2\,,
\end{equation}
where $g_2 = 16.9$ PE/electron and the charge yield $Q_y^\mathrm{X}(T_\mathcal{N})$ is taken from Ref.~\cite{XENON:2024xgd}. For PandaX-4T we use instead

\begin{equation}
    n^\mathrm{P} = N_{e^-} = T_\mathcal{N}Q_y^\mathrm{P}(T_\mathcal{N})\, ,
\end{equation}
with the charge yield $Q_y^\mathrm{P}(T_\mathcal{N})$ given in Ref.~\cite{PandaX:2024muv}. 

The overall predicted number of events in a given bin $i$  is eventually given by 
\begin{equation}
    N_i^\mathrm{X,P} = R_i^\mathrm{X,P}  + \sum_k B_i^k\,,
\label{eq:pred_n_evs}
\end{equation}
where the spectra of the background components $B_i^k$ are taken from Refs.~\cite{XENON:2024hup,PandaX:2024muv}.
Regarding the experimental data measured by the two Collaborations, $D^\mathrm{X,P}_k$, XENONnT has observed $\sum_k D^\mathrm{X}_k = 37$ events, accounting for both ionization and scintillation signals. In the case of PandaX-4T, 3 (332) events are observed for the paired (US2) signals. 
Our predictions are hence compared with the data $D^\mathrm{X,P}_k$ using 

\begin{equation}
    \chi^2_\mathrm{X,P}  = \min_{\alpha,\vec{\beta}} \left\{2\left(\sum_k N^\mathrm{X,P} _k - D^\mathrm{X,P} _k + D^{X,P}_k~\ln \left(D^\mathrm{X,P} _k/N^\mathrm{X,P}_k\right)\right) + (\alpha/\sigma_{\alpha})^2 + \sum_i (\beta_i/\sigma_{\beta_i})^2\right\}\,,
\end{equation}
where $\alpha$ is a nuisance parameter with $\sigma_\alpha = 12\%$ uncertainty accounting for the $^8$B flux prediction, and $\vec{\beta}$ and $\sigma_{\beta_i}$ are the remaining nuisance parameters and uncertainties of the experiments. All nuisance parameters are included as normalization factors on the components in Eq.~\eqref{eq:pred_n_evs}. In the case of XENONnT we include an uncertainty of 5\% on our signal prediction 
related to the fiducial volume. In addition, the background components receive the following uncertainties: a 4.8\% uncertainty for accidental coincidence (AC), a 50\% uncertainty for the neutron-related background and a 100\% uncertainty for the electron recoil (ER) background, both being subleading compared to AC. For PandaX-4T we use a 22\% uncertainty on the signal prediction due to data selection and interaction modelling. We further include uncertainties of 31\% and 23\% for the cathode and micro-discharges (MD) background components.
Finally, let us note that we also perform a combined analysis of XENONnT and PandaX-4T data. In this case, the correlated uncertainty on the neutrino flux is included only once.

\begin{figure}[!t]
\centering
\includegraphics[width=0.49\textwidth]{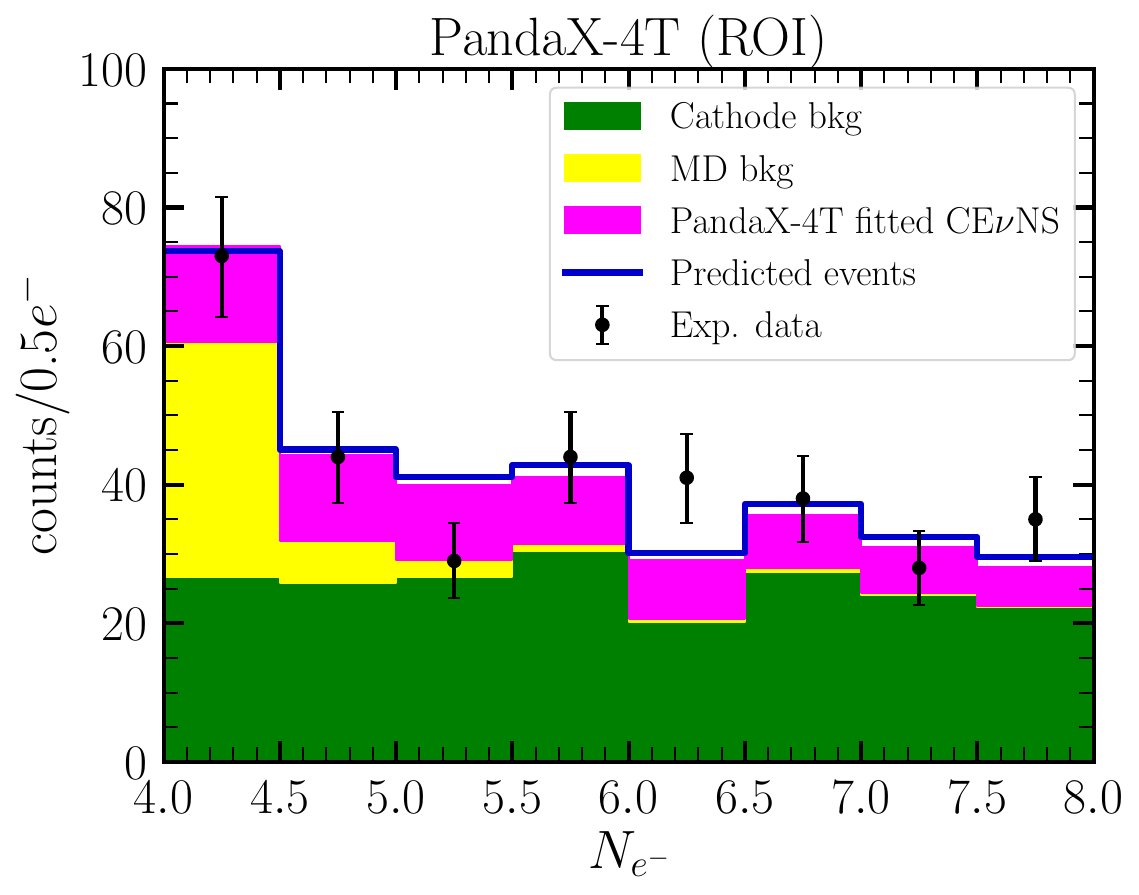}
\includegraphics[width=0.48\textwidth]{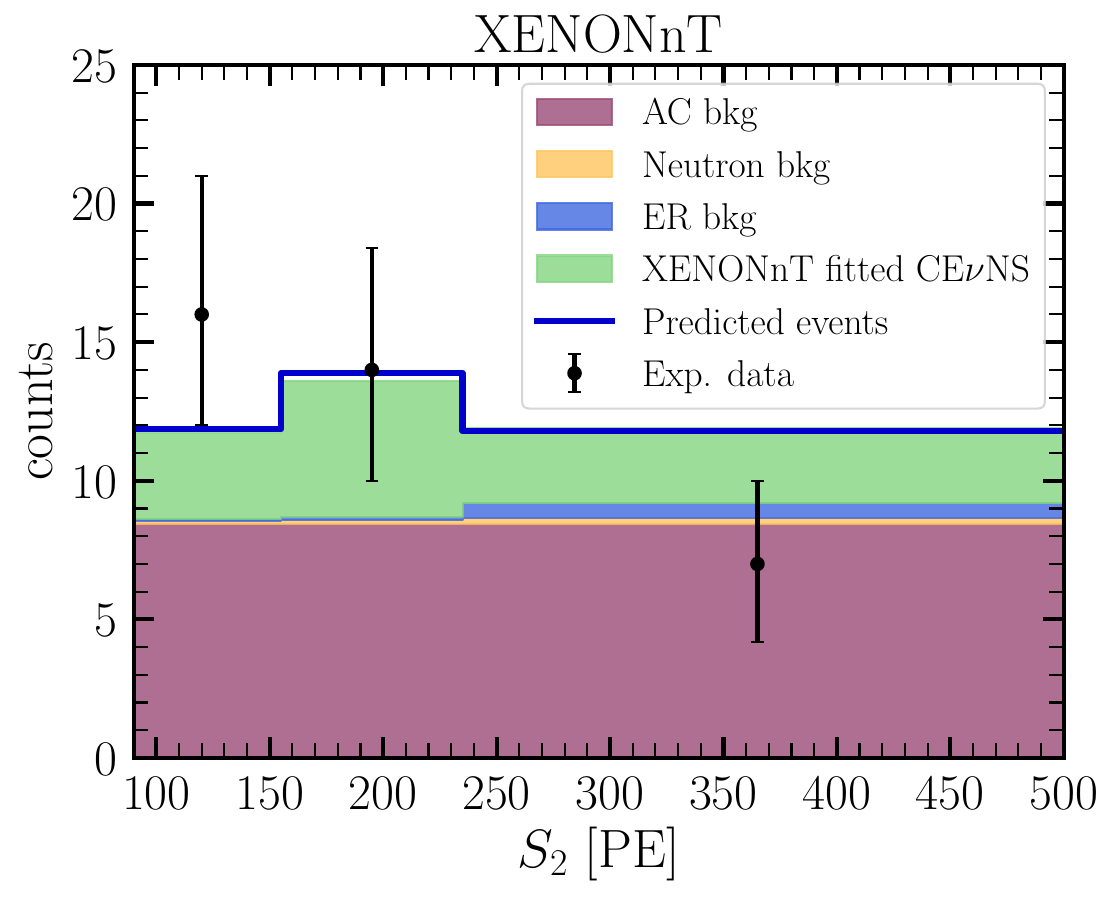}
\caption{Distribution of signal and background fitted events for PandaX-4T (left panel) and XENONnT (right panel). The colored histograms are given in the experimental papers, while the blue lines are our predictions of \cevns~plus background events for the two analyses. The measured events together with the error bars are also shown for each experiment.}
\label{fig:Nevents}
\end{figure}

In Fig.~\ref{fig:Nevents} we demonstrate the distributions of  signal and background events as a function of the number of ionized electrons, for PandaX-4T (left panel) and XENONnT (right panel). In the case of PandaX-4T, the green histogram represents the radioactivity on the cathode electrode (CE) while the micro-discharging  background is summed over the cathode background in yellow. The magenta histogram finally accounts for the \cevns~prediction plus both the CE and MD backgrounds, as given by the experimental Collaboration. Our total prediction is given as a blue line, and has to be compared to the magenta histogram. Experimental data are also shown together with their error bars.  In the case of XENONnT, the \cevns~signal is represented in light green on top of the backgrounds, indicated by light purple (AC) and light blue (electron recoil). The neutron recoil background is barely visible in the plot, but it is also included.

\section{Results}
\label{sec:results}
In this section we present the results of our analyses first concerning SM physics ($^8$B solar neutrino flux and a determination of the weak mixing angle) in Sec.~\ref{sec:weak_mix_angle} and then for new light mediators in Sec.~\ref{sec:resngi}.

\subsection{SM physics: weak mixing angle and $^8$B flux}
\label{sec:weak_mix_angle}

\begin{figure}[!t]
\centering
\includegraphics[width=0.49\textwidth]{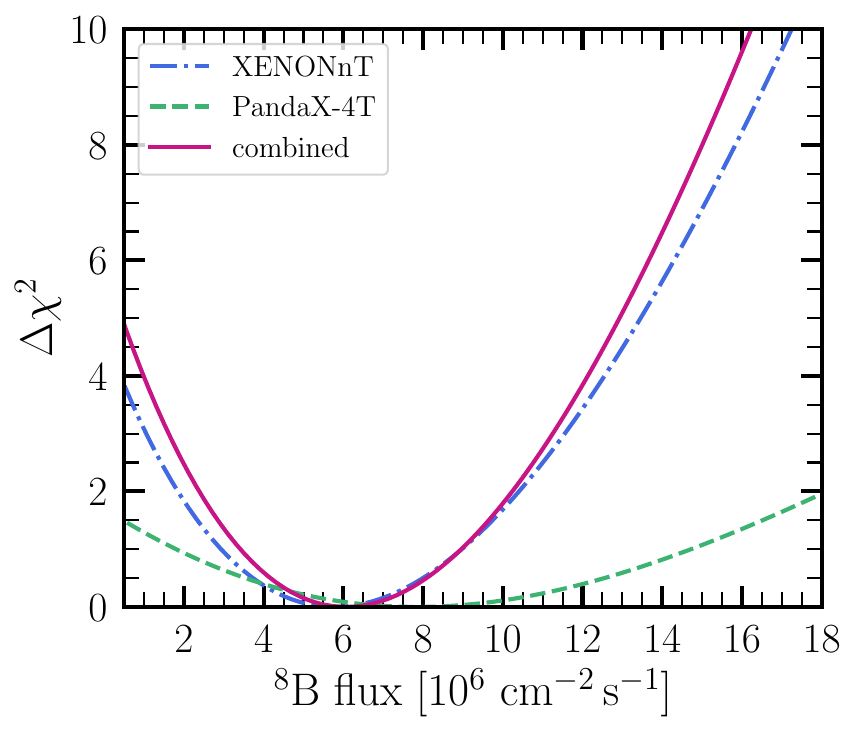}
\includegraphics[width=0.49\textwidth]{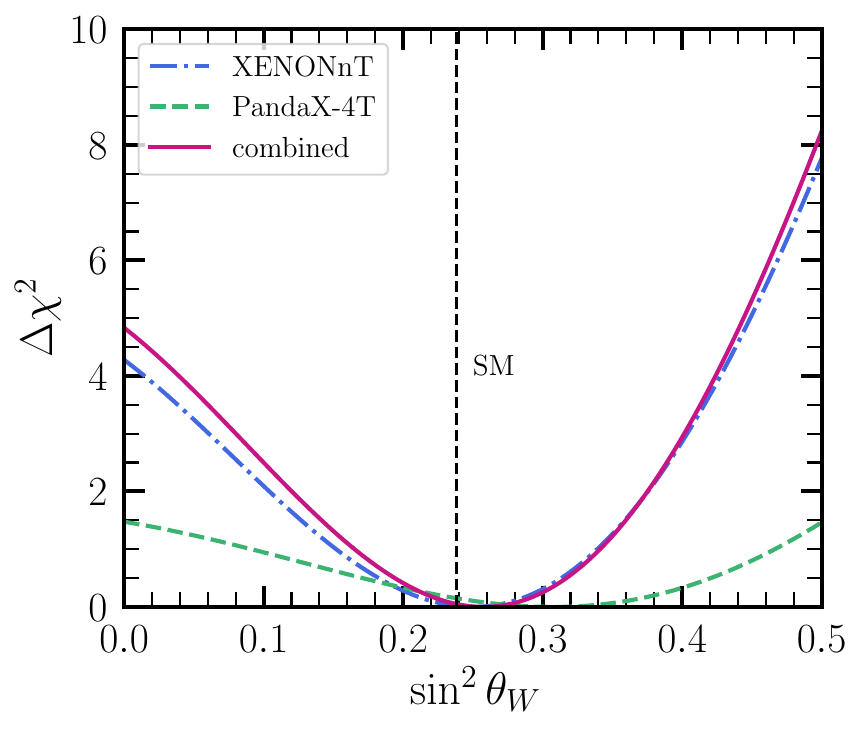}
\caption{Reduced $\chi^2$-profiles for the determination of the $^8$B solar neutrino flux (left) and of the weak mixing angle (right) for the PandaX-4T (green dashed), XENONnT (blue dot-dashed) and the combined (magenta plain) analyses. In the right plot, the vertical black dashed line indicates the SM value from the RGE running in the 
$\overline{\text{MS}}$ renormalization scheme. }
\label{fig:deltachi2}
\end{figure}

The experimental results announced by the XENONnT and PandaX-4T Collaborations allow for a measurement of the $^8$B solar neutrino flux through its \cevns-induced signal. Both Collaborations claim agreement with the standard solar model prediction and with other dedicated solar neutrino experiments, indicating a constraint of  $\Phi_\nu^\mathrm{^8B} = (4.7^{+3.6}_{-2.3}) \times 10^6~\mathrm{cm^{-2}~s^{-1}}$ at $68\%$ confidence level (CL) in the case of XENONnT and $\Phi_\nu^\mathrm{^8B} = (8.4 \pm 3.1) \times 10^6~\mathrm{cm^{-2}~s^{-1}}$ at $68\%$ CL for PandaX-4T, obtained using a combined analysis of paired and US2 data.  To test our statistical analysis, we also extract the constraints on the $^8$B solar neutrino flux for both experiments separately, and from a combined analysis. At this scope, we assume the flux-weighted \cevns~cross section as predicted in the SM, fixing for this analysis $\sin^2 \theta_W = 0.23857$. The reduced $\chi^2$-profiles are shown in the left panel of Fig.~\ref{fig:deltachi2}: the green dashed curve corresponds to  PandaX-4T, the blue dot-dashed one to XENONnT, while the magenta plain one depicts the combined result. At $1 \sigma$ CL our results read

\begin{eqnarray}
\Phi_\nu^\mathrm{^8B} &=& (7.7^{+7.0}_{-5.9}) \times 10^6 ~\mathrm{cm}^{-2}~\mathrm{s}^{-1} \quad \text{(PandaX-4T)},
\\
\Phi_\nu^\mathrm{^8B} &=& (5.7^{+3.2}_{-2.8}) \times 10^6 ~\mathrm{cm}^{-2}~\mathrm{s}^{-1} \quad \text{(XENONnT)},
\\
\Phi_\nu^\mathrm{^8B} &=& (6.1^{+2.8}_{-2.7}) \times 10^6 ~\mathrm{cm}^{-2}~\mathrm{s}^{-1} \quad \text{(combined)}.
\end{eqnarray}

Our result does not agree very well with the official result from PandaX-4T stated above.  However, since we use a reduced data set (only US2) a weaker result on the solar neutrino flux could be expected. We verified, in any case, that for our best fit value the overall number of events (we obtain 69) lies within the stated $1\sigma$ interval of the Collaboration for the US2-only analysis, which is~\cite{PandaX:2024muv}: $92\pm$34.

As anticipated in Sec.~\ref{sec:CEvNS}, one relevant SM parameter entering the \cevns~cross section is the weak mixing angle, sin$^2 \theta_W$. The observation of \cevns~data at DM direct detection experiments allow to determine this parameter at low energy, i.e., at a renormalization scale $\mu \simeq \mathcal{O}(10)$ MeV, corresponding to the typical momentum transfer exchanged in the process. A variation in the value of sin$^2 \theta_W$ would affect the overall normalization of the \cevns~event rates. We perform a statistical analysis, this time fixing $\Phi_\nu^\mathrm{^8B} = 5.46 \times 10^6 ~\mathrm{cm^{-2}~s^{-1}}$ (but adding the associated 12\% uncertainty), and we extract the following best fit values and $1 \sigma$ uncertainties 

\begin{eqnarray}
     \sin^2 \theta_W &=& 0.30^{+0.16}_{-0.21} \quad \text{(PandaX-4T)},\\
     \sin^2 \theta_W &=& 0.25^{+0.09}_{-0.10} \quad \text{(XENONnT)},\\
     \sin^2 \theta_W &=& 0.26^{+0.08}_{-0.09} \quad \text{(combined)}. 
\end{eqnarray}

\begin{figure}[!t]
\centering
\includegraphics[width=0.65\textwidth]{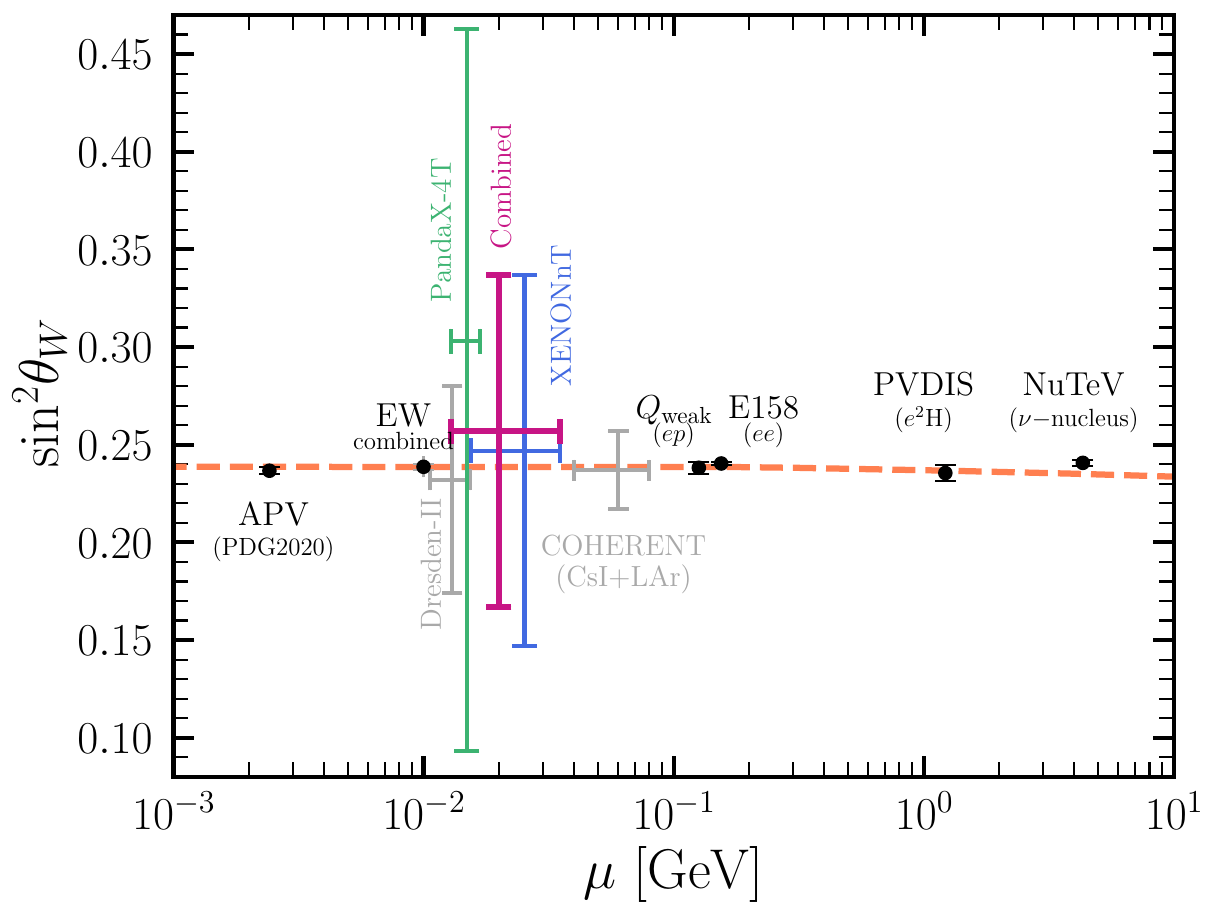}
\caption{Weak mixing angle running in the SM for the  $\overline{\text{MS}}$ renormalization scheme (coral dashed line) as a function of the renormalization scale. Our  $1 \sigma$ determinations are shown in green for PandaX-4T, blue for XENONnT and magenta for the combined analysis. Measurements from other experiments~\cite{Majumdar:2022nby,DeRomeri:2022twg,ParticleDataGroup:2024cfk,Qweak:2018tjf,SLACE158:2005uay,PVDIS:2014cmd,NuTeV:2001whx,AtzoriCorona:2024vhj} are also shown for comparison.}
\label{fig:res_sw2}
\end{figure}

Figure~\ref{fig:deltachi2} (right) shows the reduced $\chi^2$-profiles for the determination of the weak mixing angle, for the two separate data sets and for the combined analysis.
Moreover, we show in Fig.~\ref{fig:res_sw2} the best fit values together with the $1 \sigma$ error bars, as a function of the renormalization scale. For comparison, the plot additionally shows the RGE evolution in the SM (coral dashed line), calculated in the 
$\overline{\text{MS}}$ renormalization scheme~\cite{Erler:2004in} as well as other existing constraints at different energy scales~\cite{Wood:1997zq,Qweak:2018tjf,SLACE158:2005uay,PVDIS:2014cmd,NuTeV:2001whx}. 
Let us note that a determination of sin$^2 \theta_W$ from recent direct detection data was already performed in~\cite{Maity:2024aji} leading to similar results in the case of XENONnT data. However, in the case of PandaX-4T our result differs from Ref.~\cite{Maity:2024aji}. This might be due to the fact that our Eq.~\eqref{eq:T2n_varchange} does not agree with Eq.~(7) in Ref.~\cite{Maity:2024aji}, since the charge yield $Q_y$ also depends on the nuclear recoil energy and hence $dT_\mathcal{N}/dn$ is not $1/Q_y$ as assumed in this reference.

Complementarity with other \cevns~measurements is particularly evident, for instance those from COHERENT CsI and liquid argon data~\cite{DeRomeri:2022twg}, from the Dresden-II reactor experiment~\cite{AristizabalSierra:2022axl,Majumdar:2022nby}, represented in gray, and from a combination of different electroweak measurements~\cite{AtzoriCorona:2024vhj}. Even though not shown in the plot to avoid overcrowding the figure, other low-energy measurements have been extracted from spallation source and reactor \cevns~data~\cite{Cadeddu:2020lky,AtzoriCorona:2022qrf,AtzoriCorona:2023ktl}, also in combination with data from atomic parity violation and parity-violating electron scattering on different nuclei~\cite{Cadeddu:2021ijh,Cadeddu:2024baq}. Sensitivities from elastic scattering off atomic electrons at IsoDAR~\cite{Alonso:2021kyu}, neutrino-electron scattering events at SBND~\cite{Alves:2024twb} and DUNE~\cite{deGouvea:2019wav} further complete the picture of low-energy sin$^2 \theta_W$ determinations.
DM direct detection facilities, despite their different primary scientific goal, can still provide valuable information on the value of the neutrino neutral-current interaction at low energy. While the current error bars of the measurements at DM direct detection experiments are still large compared to other determinations at higher energy scales, these novel measurements lie in a part of parameter space previously unexplored. Forthcoming data and improved statistics at DM facilities will allow to extract the value of sin$^2 \theta_W$ at $\mu \simeq \mathcal{O}(10)$ MeV with higher precision, in full complementarity with dedicated \cevns~experiments.

\subsection{Neutrino generalized interactions}
\label{sec:resngi}
Next we discuss the constraints obtained for new neutrino interactions in the presence of different light mediators, as defined in Sec.~\ref{subsec:CEvNS-NGI}. In these analyses, we have kept the value of the weak mixing angle fixed at its SM value, $\sin^2\theta_W=0.23857$ and the normalization of the $^8$B solar neutrino flux $\Phi_\nu^\mathrm{^8B} = 5.46 \times 10^6~\mathrm{cm^{-2}~s^{-1}}$. 

\begin{figure}[!htb]
\centering
\includegraphics[width=0.49\textwidth]{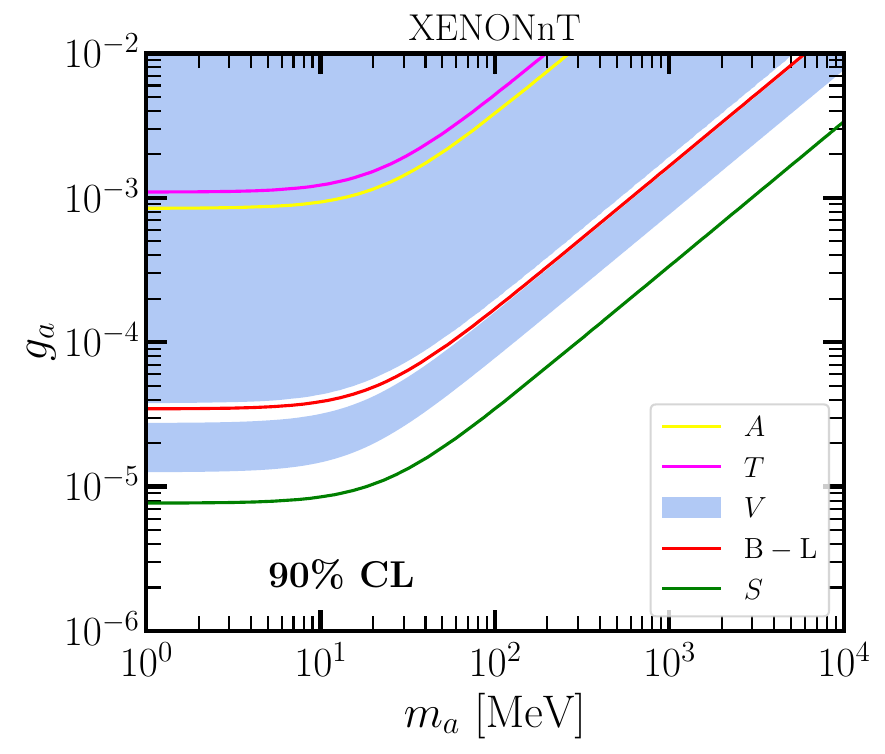}
\includegraphics[width=0.49\textwidth]{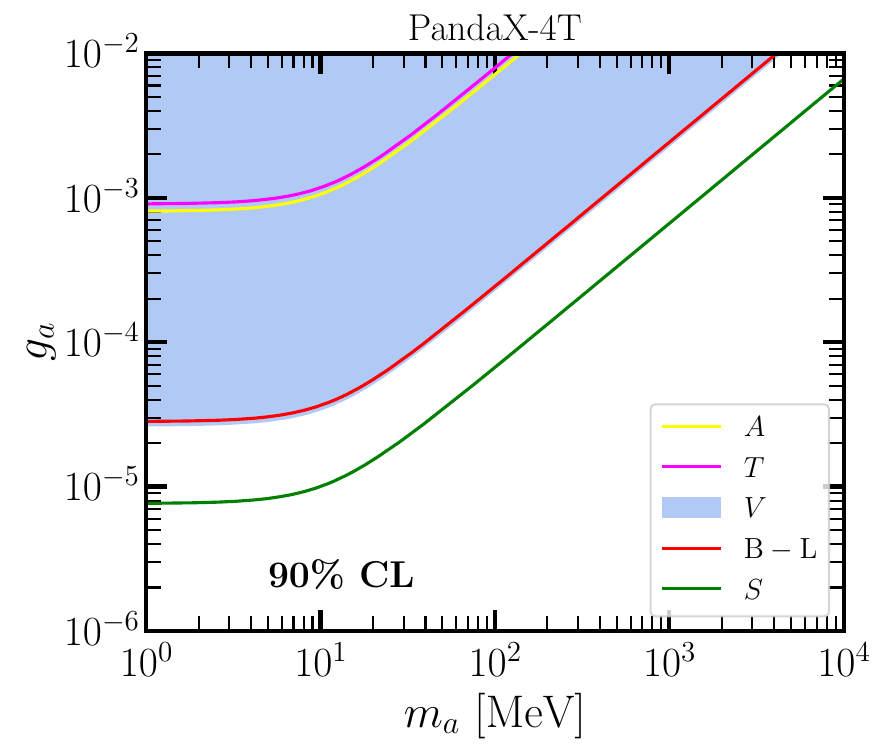}
\\
\includegraphics[width=0.49\textwidth]{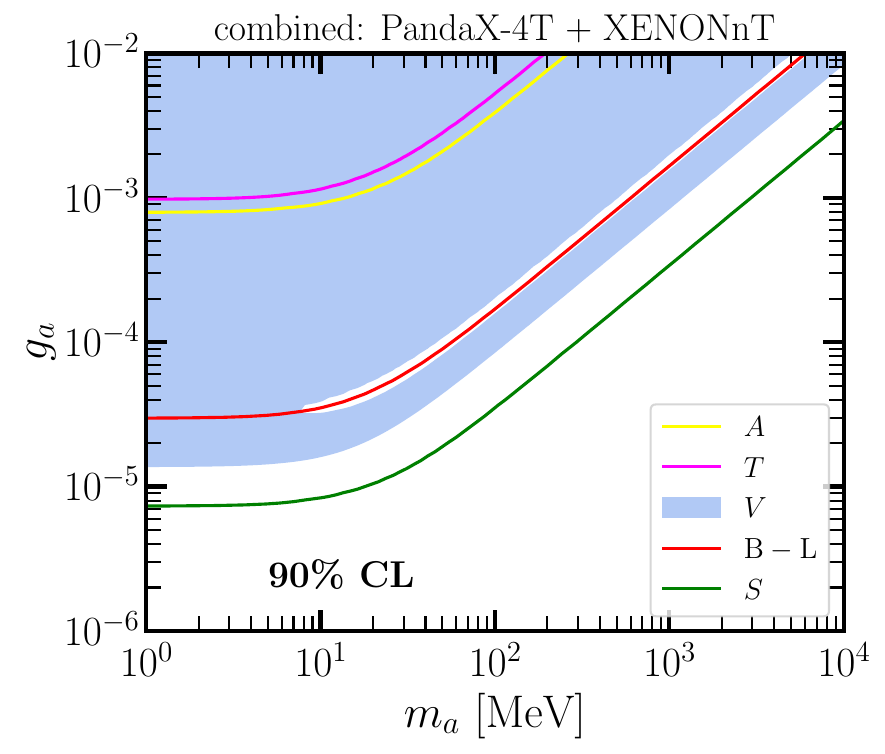}
\caption{The 90\% CL exclusion limits for new neutrino generalized interactions with light mediators obtained from the analysis of XENONnT data (upper left), PandaX-4T data (upper right), and from the combined analysis (lower panel).}
\label{fig:res_ngi}
\end{figure}

The results are shown in Fig.~\ref{fig:res_ngi}, where we show the contours at 90\% CL for XENONnT (upper left panel), PandaX-4T (upper right panel), and for the combined analysis (lower panel). The magenta, gold, red and green lines correspond to the analyses of tensor, axial, vector B-L, and scalar interactions as introduced in Sec.~\ref{subsec:CEvNS-NGI}. The light blue shaded region denotes the excluded region in the case of universal vector interactions. We chose this format to highlight the fact that a region in the form of a thin band remains allowed in this scenario in the case of the XENONnT and the combined analysis. This degeneracy appears due to a destructive interference between the SM and the new vector couplings in the weak nuclear charge.  In the PandaX-4T analysis we expect this degeneracy to appear below the currently excluded region once more statistics becomes available. Moreover, this cancellation can not occur in the case of the B-L model, due to the particle charges under the $U(1)_\mathrm{B-L}$ symmetry. 
As expected, the bounds on the spin-dependent axial and tensor mediators are much weaker than those for scalar and vector mediators.

\begin{figure}[t]
\centering
\includegraphics[width=0.49\textwidth]{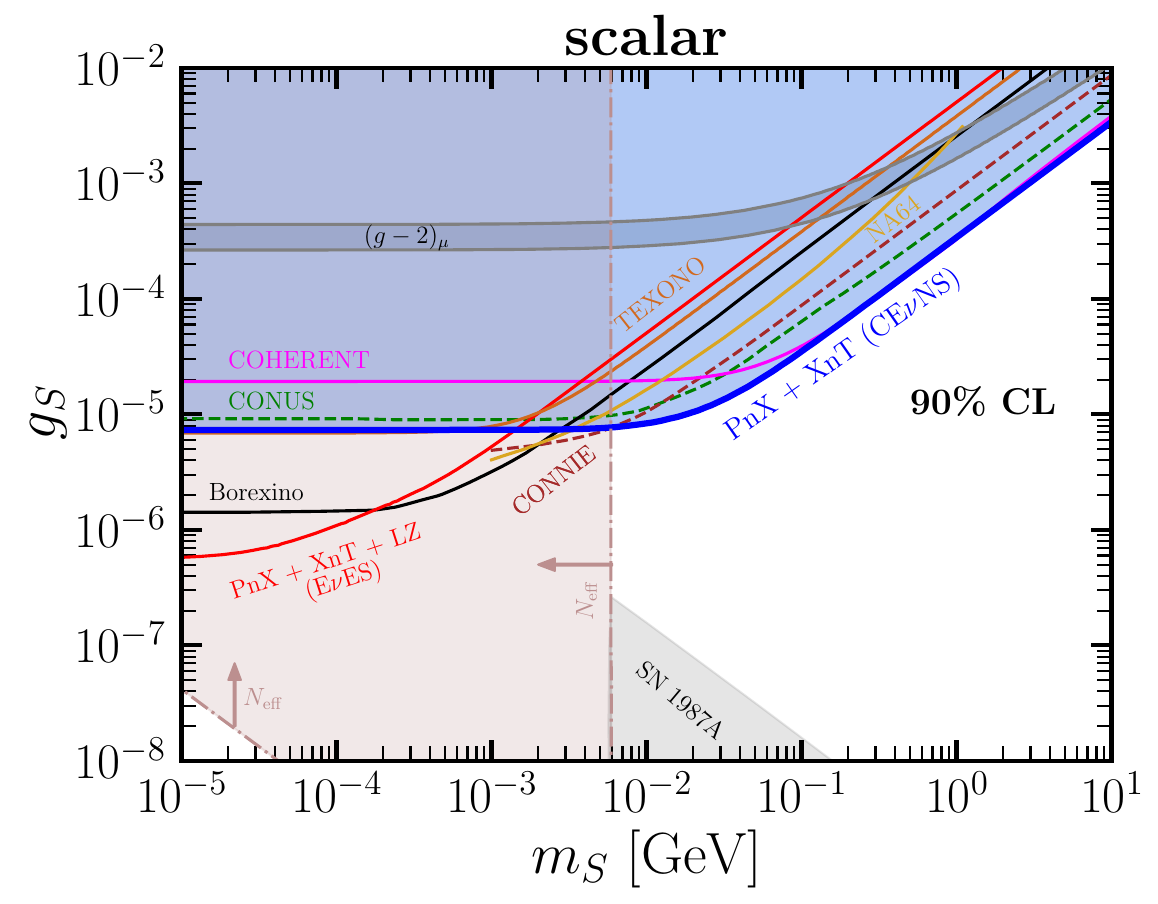}
\includegraphics[width=0.49\textwidth]{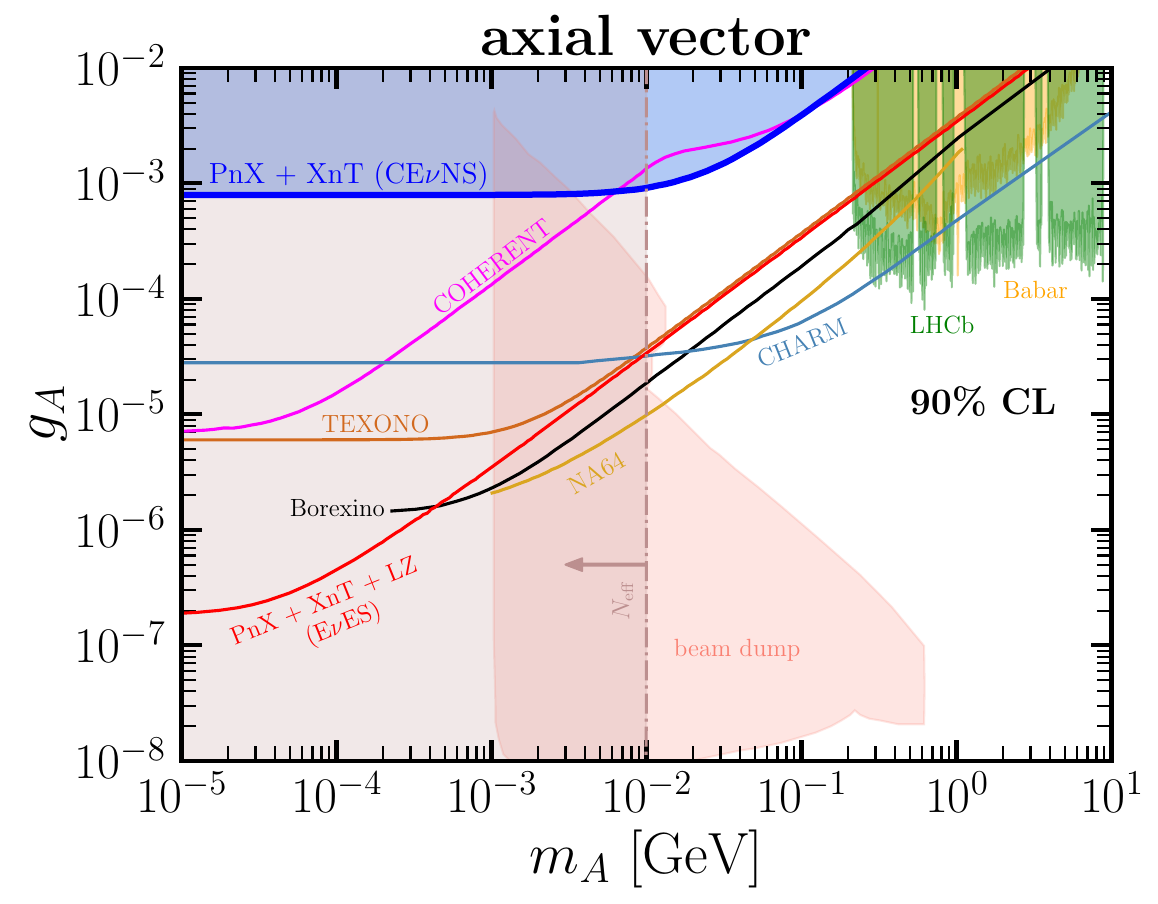}
\\
\includegraphics[width=0.49\textwidth]{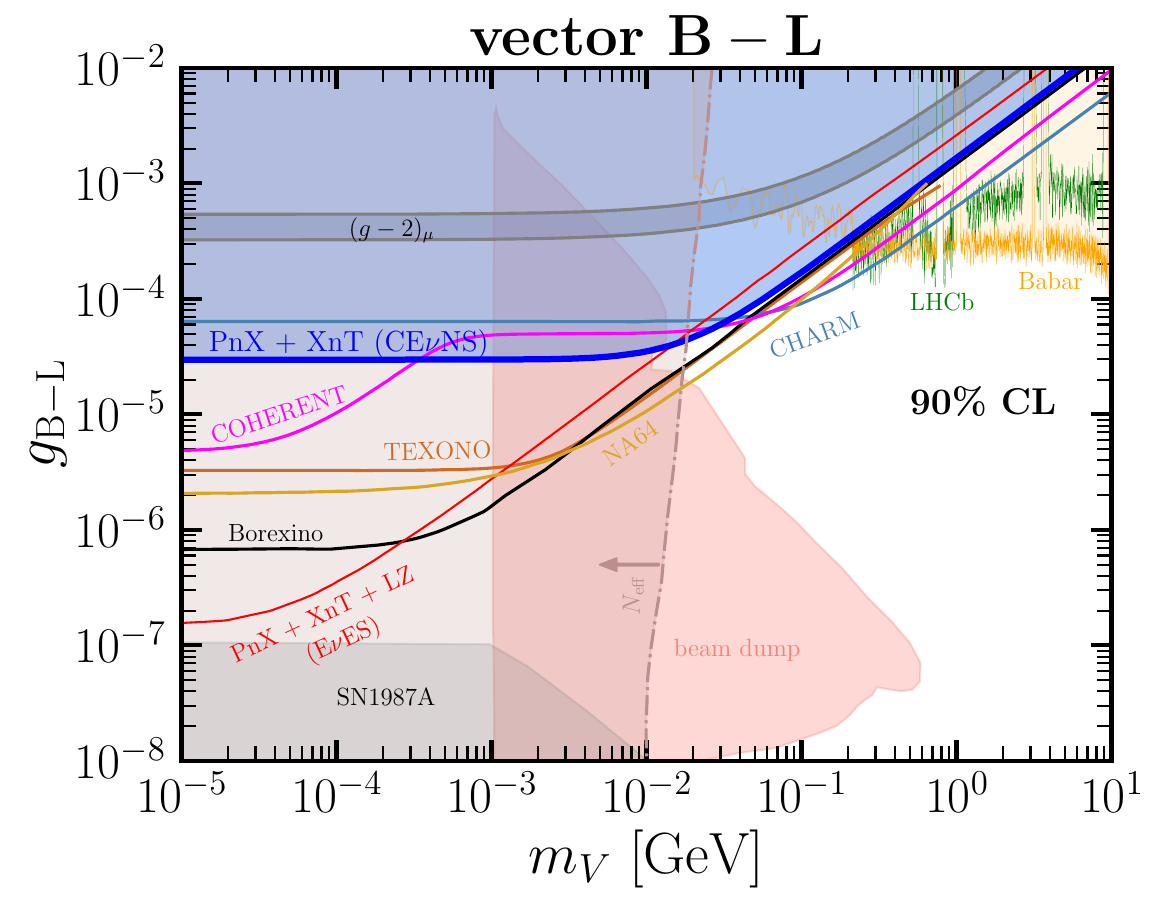}
\includegraphics[width=0.49\textwidth]{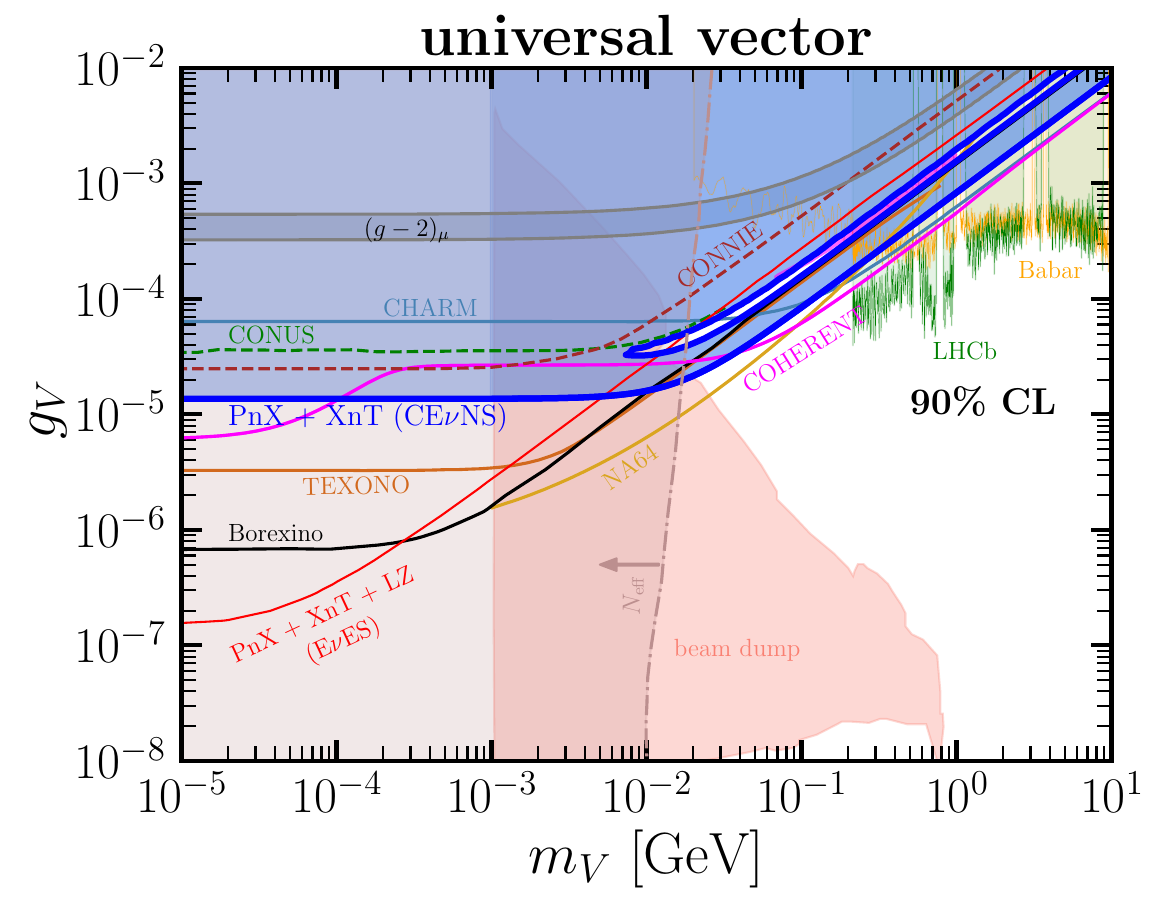}
\caption{The 90\% CL excluded regions for some interactions obtained from the combined analysis of PandaX-4T and XENONnT data (blue):  scalar (upper, left), axial-vector (upper, right), vector  B-L (lower, left) and universal vector (lower, right). Existing bounds from other searches are also shown for comparison.}
\label{fig:res_ngi_comparison}
\end{figure}

In order to put our results into context, in Fig.~\ref{fig:res_ngi_comparison} we compare our bounds with other existing constraints on the same types of interactions. We show our constraints obtained from the combined analysis of XENONnT + PandaX-4T (in blue, at $90\%$ CL) for the scalar, vector universal, vector B-L and axial-vector interactions. Additionally, we show existing limits from other \cevns~data, in particular COHERENT~\cite{DeRomeri:2022twg,AtzoriCorona:2022moj}, CONUS~\cite{CONUS:2021dwh,Lindner:2024eng}, and CONNIE~\cite{CONNIE:2019xid,CONNIE:2024pwt}; from elastic neutrino-electron scattering data at BOREXINO~\cite{Coloma:2022avw}, CHARM-II~\cite{Bauer:2018onh} and TEXONO~\cite{TEXONO:2009knm,Bauer:2018onh}; from a combined analysis of PandaX-4T, XENONnT and LZ electron recoil data~\cite{A:2022acy,DeRomeri:2024dbv}; from beam-dump and fixed-target experiments
(including E141~\cite{Riordan:1987aw},
E137~\cite{Bjorken:1988as},
E774~\cite{Bross:1989mp},
KEK~\cite{Konaka:1986cb},
Orsay~\cite{Davier:1989wz,Bjorken:2009mm,Andreas:2012mt},
$\nu$-CAL~I~\cite{Blumlein:1990ay,Blumlein:1991xh,Blumlein:2011mv,Blumlein:2013cua},
CHARM~\cite{CHARM:1985anb,Gninenko:2012eq},
NOMAD~\cite{NOMAD:2001eyx},
PS191~\cite{Bernardi:1985ny,Gninenko:2011uv}, A1~\cite{Merkel:2014avp} and APEX~\cite{APEX:2011dww}); from
colliders
(BaBar~\cite{BaBar:2014zli,BaBar:2017tiz} and LHCb~\cite{LHCb:2017trq}); from NA64~\cite{NA64:2021xzo,NA64:2022yly,NA64:2023wbi}. To recast some of the bounds between the different interactions we used the DarkCast package~\cite{Ilten:2018crw,Baruch:2022esd}. Note also that some of the bounds are a novel result in this work: we have recomputed the COHERENT axial-vector bound following~\cite{DeRomeri:2022twg}, however including only the analysis of CsI, since the $^{40}$Ar nucleus is even. The combined XENONnT + PandaX-4T + LZ \eves~bounds for the scalar, the universal vector and the axial-vector, not shown in~\cite{DeRomeri:2024dbv}, have also been computed specifically for this work. Similarly, the TEXONO bound on the scalar interaction is a new result. 
Finally, astrophysical and
cosmological bounds are also particularly relevant for low-mass mediators. Even though strongly
model-dependent and thus requiring a tailored analysis, we indicate with $N_\mathrm{eff}$ the regions potentially in conflict with BBN~\cite{Esseili:2023ldf,Li:2023puz,Ghosh:2024cxi} and CMB~\cite{PhysRevD.110.075032} and with SN1987A those in conflict supernova~\cite{Heurtier:2016otg,Chang:2016ntp,Croon:2020lrf,Caputo:2021rux,Caputo:2022rca} data.

As can be seen, for the case of scalar-mediated processes, the constraints extracted in this work dominate for $6<m_S<150$~MeV, improving previous results from  dedicated \cevns~experiments such as CONUS, CONNIE and COHERENT, while for $m_S>150$~MeV the sensitivity becomes similar to COHERENT. It is also noteworthy that the present sensitivities are the leading ones among the \cevns-based measurements in the region that is not in conflict with astrophysics. For the case of axial-vector interactions, as previously noted, the nuclear spin-suppression leads to poor sensitivities compared to e.g., the \eves-induced constraints from TEXONO, CHARM and Borexino as well as to those coming from a combined analysis of electron recoils at PandaX-4T, XENONnT and LZ. However, if the axial mediator coupling to quarks is assumed to be different from the coupling to leptons, then the present results ---although very weak--- are dominating in the mass range $6<m_A<80$~MeV, while for larger masses they eventually become similar to COHERENT. 
Finally, focusing on the B-L and universal vector mediator models, the
present sensitivities are not improving upon existing constraints, though being almost competitive in some regions of the parameter space. When compared to other CEνNS constraints like COHERENT,
XENONnT and PandaX-4T offer a slight improvement in the range $0.1 < m_V < 30$~MeV ($0.3 < m_V < 40$~MeV) in the B-L
(universal vector) case, in a region that is however in tension with astrophysical observations.

\section{Conclusions and outlook}
\label{sec:concl}

Indications of $^8$B solar neutrinos inducing \cevns~at DM direct detection experiments have recently been reported.
Motivated by these results, we have analyzed the first \cevns~data collected by the PandaX-4T and XENONnT experiments. 
We have performed SM and new physics analyses showing that these data can be used to determine the weak mixing angle and to place a bound on the solar $^8$B neutrino flux. The obtained constraints on the weak mixing angle lie at a low energy scale, falling between the Dresden-II and COHERENT measurements. In addition, we have placed bounds on the mass and coupling of several light mediators, focusing on those with scalar, vector (universal and B-L), axial-vector, and tensor interactions. We have found that even with these first data we can place competitive bounds in some regions of parameter space, specially in the case of scalar and vector interactions, complementing other experimental probes including neutrino scattering data, beam dump and collider searches. In this paper we have focused on universal interactions, i.e., assuming equal couplings for all neutrino species, charged leptons and quarks. In these scenarios the bound is independent of the composition of the neutrino flux at the detector. In a future work we plan to extend the analysis including searches with non-universal and flavor-dependent couplings. More data is expected to be released in the future, from the DM experiments analyzed in this paper and also from the LZ experiment, which will allow us to further improve the bounds obtained here.

\section*{Acknowledgments}
The authors thank Riccardo Biondi (from the XENONnT Collaboration), Matthew Szydagis and Vetri Velan (from the NEST Collaboration) for helpful discussions. We also acknowledge useful discussions regarding Darkcast with José Zurita and about NA64 with Laura Molina Bueno as well as Sk Jeesun for sharing information regarding constraints from $N_\mathrm{eff}$.
The work of VDR is supported by the CIDEXG/2022/20 grant funded by Generalitat Valenciana. VDR and DKP acknowledge financial support from the Spanish grants CNS2023-144124 (MCIN/AEI/10.13039/501100011033 and “Next Generation EU”/PRTR), PID2023-147306NB-I00 and CEX2023-001292-S (MCIU/AEI/ 10.13039/501100011033).

\bibliographystyle{utphys}
\bibliography{bibliography}  

\providecommand{\href}[2]{#2}\begingroup\raggedright\begin{thebibliography}{100}

\bibitem{Abdullah:2022zue}
M.~Abdullah {\em et~al.}, ``{Coherent elastic neutrino-nucleus scattering:
  Terrestrial and astrophysical applications},''
  \href{http://arxiv.org/abs/2203.07361}{{\ttfamily arXiv:2203.07361
  [hep-ph]}}.

\bibitem{Freedman:1973yd}
D.~Z. Freedman, ``{Coherent Neutrino Nucleus Scattering as a Probe of the Weak
  Neutral Current},'' \href{http://dx.doi.org/10.1103/PhysRevD.9.1389}{{\em
  Phys. Rev. D} {\bfseries 9} (1974) 1389--1392}.

\bibitem{Drukier:1984vhf}
A.~Drukier and L.~Stodolsky, ``{Principles and Applications of a Neutral
  Current Detector for Neutrino Physics and Astronomy},''
  \href{http://dx.doi.org/10.1103/PhysRevD.30.2295}{{\em Phys. Rev. D}
  {\bfseries 30} (1984) 2295}.

\bibitem{COHERENT:2017ipa}
{\bfseries COHERENT} Collaboration, D.~Akimov {\em et~al.}, ``{Observation of
  Coherent Elastic Neutrino-Nucleus Scattering},''
  \href{http://dx.doi.org/10.1126/science.aao0990}{{\em Science} {\bfseries
  357} (2017) 1123--1126}, \href{http://arxiv.org/abs/1708.01294}{{\ttfamily
  arXiv:1708.01294 [nucl-ex]}}.

\bibitem{COHERENT:2020iec}
{\bfseries COHERENT} Collaboration, D.~Akimov {\em et~al.}, ``{First
  Measurement of Coherent Elastic Neutrino-Nucleus Scattering on Argon},''
  \href{http://dx.doi.org/10.1103/PhysRevLett.126.012002}{{\em Phys. Rev.
  Lett.} {\bfseries 126} no.~1, (2021) 012002},
  \href{http://arxiv.org/abs/2003.10630}{{\ttfamily arXiv:2003.10630
  [nucl-ex]}}.

\bibitem{COHERENT:2021xmm}
{\bfseries COHERENT} Collaboration, D.~Akimov {\em et~al.}, ``{Measurement of
  the Coherent Elastic Neutrino-Nucleus Scattering Cross Section on CsI by
  COHERENT},'' \href{http://dx.doi.org/10.1103/PhysRevLett.129.081801}{{\em
  Phys. Rev. Lett.} {\bfseries 129} (2022) 081801},
  \href{http://arxiv.org/abs/2110.07730}{{\ttfamily arXiv:2110.07730
  [hep-ex]}}.

\bibitem{Adamski:2024yqt}
S.~Adamski {\em et~al.}, ``{First detection of coherent elastic
  neutrino-nucleus scattering on germanium},''
  \href{http://arxiv.org/abs/2406.13806}{{\ttfamily arXiv:2406.13806
  [hep-ex]}}.

\bibitem{Colaresi:2022obx}
J.~Colaresi, J.~I. Collar, T.~W. Hossbach, C.~M. Lewis, and K.~M. Yocum,
  ``{Measurement of Coherent Elastic Neutrino-Nucleus Scattering from Reactor
  Antineutrinos},''
  \href{http://dx.doi.org/10.1103/PhysRevLett.129.211802}{{\em Phys. Rev.
  Lett.} {\bfseries 129} no.~21, (2022) 211802},
  \href{http://arxiv.org/abs/2202.09672}{{\ttfamily arXiv:2202.09672
  [hep-ex]}}.

\bibitem{Bertone:2016nfn}
G.~Bertone and D.~Hooper, ``{History of dark matter},''
  \href{http://dx.doi.org/10.1103/RevModPhys.90.045002}{{\em Rev. Mod. Phys.}
  {\bfseries 90} no.~4, (2018) 045002},
  \href{http://arxiv.org/abs/1605.04909}{{\ttfamily arXiv:1605.04909
  [astro-ph.CO]}}.

\bibitem{Goodman:1984dc}
M.~W. Goodman and E.~Witten, ``{Detectability of Certain Dark Matter
  Candidates},'' \href{http://dx.doi.org/10.1103/PhysRevD.31.3059}{{\em Phys.
  Rev. D} {\bfseries 31} (1985) 3059}.

\bibitem{Schumann:2019eaa}
M.~Schumann, ``{Direct Detection of WIMP Dark Matter: Concepts and Status},''
  \href{http://dx.doi.org/10.1088/1361-6471/ab2ea5}{{\em J. Phys. G} {\bfseries
  46} no.~10, (2019) 103003}, \href{http://arxiv.org/abs/1903.03026}{{\ttfamily
  arXiv:1903.03026 [astro-ph.CO]}}.

\bibitem{Billard:2021uyg}
J.~Billard {\em et~al.}, ``{Direct detection of dark matter\textemdash{}APPEC
  committee report*},'' \href{http://dx.doi.org/10.1088/1361-6633/ac5754}{{\em
  Rept. Prog. Phys.} {\bfseries 85} no.~5, (2022) 056201},
  \href{http://arxiv.org/abs/2104.07634}{{\ttfamily arXiv:2104.07634
  [hep-ex]}}.

\bibitem{XENON:2023cxc}
{\bfseries XENON} Collaboration, E.~Aprile {\em et~al.}, ``{First Dark Matter
  Search with Nuclear Recoils from the XENONnT Experiment},''
  \href{http://dx.doi.org/10.1103/PhysRevLett.131.041003}{{\em Phys. Rev.
  Lett.} {\bfseries 131} no.~4, (2023) 041003},
  \href{http://arxiv.org/abs/2303.14729}{{\ttfamily arXiv:2303.14729
  [hep-ex]}}.

\bibitem{LZ:2022lsv}
{\bfseries LZ} Collaboration, J.~Aalbers {\em et~al.}, ``{First Dark Matter
  Search Results from the LUX-ZEPLIN (LZ) Experiment},''
  \href{http://dx.doi.org/10.1103/PhysRevLett.131.041002}{{\em Phys. Rev.
  Lett.} {\bfseries 131} no.~4, (2023) 041002},
  \href{http://arxiv.org/abs/2207.03764}{{\ttfamily arXiv:2207.03764
  [hep-ex]}}.

\bibitem{PandaX-4T:2021bab}
{\bfseries PandaX-4T} Collaboration, Y.~Meng {\em et~al.}, ``{Dark Matter
  Search Results from the PandaX-4T Commissioning Run},''
  \href{http://dx.doi.org/10.1103/PhysRevLett.127.261802}{{\em Phys. Rev.
  Lett.} {\bfseries 127} no.~26, (2021) 261802},
  \href{http://arxiv.org/abs/2107.13438}{{\ttfamily arXiv:2107.13438
  [hep-ex]}}.

\bibitem{Monroe:2007xp}
J.~Monroe and P.~Fisher, ``{Neutrino Backgrounds to Dark Matter Searches},''
  \href{http://dx.doi.org/10.1103/PhysRevD.76.033007}{{\em Phys. Rev. D}
  {\bfseries 76} (2007) 033007},
  \href{http://arxiv.org/abs/0706.3019}{{\ttfamily arXiv:0706.3019
  [astro-ph]}}.

\bibitem{Vergados:2008jp}
J.~D. Vergados and H.~Ejiri, ``{Can Solar Neutrinos be a Serious Background in
  Direct Dark Matter Searches?},''
  \href{http://dx.doi.org/10.1016/j.nuclphysb.2008.06.004}{{\em Nucl. Phys. B}
  {\bfseries 804} (2008) 144--159},
  \href{http://arxiv.org/abs/0805.2583}{{\ttfamily arXiv:0805.2583 [hep-ph]}}.

\bibitem{Strigari:2009bq}
L.~E. Strigari, ``{Neutrino Coherent Scattering Rates at Direct Dark Matter
  Detectors},'' \href{http://dx.doi.org/10.1088/1367-2630/11/10/105011}{{\em
  New J. Phys.} {\bfseries 11} (2009) 105011},
  \href{http://arxiv.org/abs/0903.3630}{{\ttfamily arXiv:0903.3630
  [astro-ph.CO]}}.

\bibitem{Billard:2013qya}
J.~Billard, L.~Strigari, and E.~Figueroa-Feliciano, ``{Implication of neutrino
  backgrounds on the reach of next generation dark matter direct detection
  experiments},'' \href{http://dx.doi.org/10.1103/PhysRevD.89.023524}{{\em
  Phys. Rev. D} {\bfseries 89} no.~2, (2014) 023524},
  \href{http://arxiv.org/abs/1307.5458}{{\ttfamily arXiv:1307.5458 [hep-ph]}}.

\bibitem{AristizabalSierra:2024smb}
D.~Aristizabal~Sierra, V.~De~Romeri, and C.~A. Ternes, ``{Reactor neutrino
  background in next-generation dark matter detectors},''
  \href{http://dx.doi.org/10.1103/PhysRevD.109.115026}{{\em Phys. Rev. D}
  {\bfseries 109} no.~11, (2024) 115026},
  \href{http://arxiv.org/abs/2402.06416}{{\ttfamily arXiv:2402.06416
  [hep-ph]}}.

\bibitem{OHare:2021utq}
C.~A.~J. O'Hare, ``{New Definition of the Neutrino Floor for Direct Dark Matter
  Searches},'' \href{http://dx.doi.org/10.1103/PhysRevLett.127.251802}{{\em
  Phys. Rev. Lett.} {\bfseries 127} no.~25, (2021) 251802},
  \href{http://arxiv.org/abs/2109.03116}{{\ttfamily arXiv:2109.03116
  [hep-ph]}}.

\bibitem{Harnik:2012ni}
R.~Harnik, J.~Kopp, and P.~A.~N. Machado, ``{Exploring nu Signals in Dark
  Matter Detectors},''
  \href{http://dx.doi.org/10.1088/1475-7516/2012/07/026}{{\em JCAP} {\bfseries
  07} (2012) 026}, \href{http://arxiv.org/abs/1202.6073}{{\ttfamily
  arXiv:1202.6073 [hep-ph]}}.

\bibitem{AtzoriCorona:2022jeb}
M.~Atzori~Corona, W.~M. Bonivento, M.~Cadeddu, N.~Cargioli, and F.~Dordei,
  ``{New constraint on neutrino magnetic moment and neutrino millicharge from
  LUX-ZEPLIN dark matter search results},''
  \href{http://dx.doi.org/10.1103/PhysRevD.107.053001}{{\em Phys. Rev. D}
  {\bfseries 107} no.~5, (2023) 053001},
  \href{http://arxiv.org/abs/2207.05036}{{\ttfamily arXiv:2207.05036
  [hep-ph]}}.

\bibitem{deGouvea:2021ymm}
A.~de~Gouv\^ea, E.~McGinness, I.~Martinez-Soler, and Y.~F. Perez-Gonzalez,
  ``{pp solar neutrinos at DARWIN},''
  \href{http://dx.doi.org/10.1103/PhysRevD.106.096017}{{\em Phys. Rev. D}
  {\bfseries 106} no.~9, (2022) 096017},
  \href{http://arxiv.org/abs/2111.02421}{{\ttfamily arXiv:2111.02421
  [hep-ph]}}.

\bibitem{Giunti:2023yha}
C.~Giunti and C.~A. Ternes, ``{Testing neutrino electromagnetic properties at
  current and future dark matter experiments},''
  \href{http://dx.doi.org/10.1103/PhysRevD.108.095044}{{\em Phys. Rev. D}
  {\bfseries 108} no.~9, (2023) 095044},
  \href{http://arxiv.org/abs/2309.17380}{{\ttfamily arXiv:2309.17380
  [hep-ph]}}.

\bibitem{Cerdeno:2016sfi}
D.~G. Cerde\~no, M.~Fairbairn, T.~Jubb, P.~A.~N. Machado, A.~C. Vincent, and
  C.~B\oe{}hm, ``{Physics from solar neutrinos in dark matter direct detection
  experiments},'' \href{http://dx.doi.org/10.1007/JHEP09(2016)048}{{\em JHEP}
  {\bfseries 05} (2016) 118}, \href{http://arxiv.org/abs/1604.01025}{{\ttfamily
  arXiv:1604.01025 [hep-ph]}}. [Erratum: JHEP 09, 048 (2016)].

\bibitem{Dutta:2017nht}
B.~Dutta, S.~Liao, L.~E. Strigari, and J.~W. Walker, ``{Non-standard
  interactions of solar neutrinos in dark matter experiments},''
  \href{http://dx.doi.org/10.1016/j.physletb.2017.08.031}{{\em Phys. Lett. B}
  {\bfseries 773} (2017) 242--246},
  \href{http://arxiv.org/abs/1705.00661}{{\ttfamily arXiv:1705.00661
  [hep-ph]}}.

\bibitem{Gelmini:2018gqa}
G.~B. Gelmini, V.~Takhistov, and S.~J. Witte, ``{Geoneutrinos in Large Direct
  Detection Experiments},''
  \href{http://dx.doi.org/10.1103/PhysRevD.99.093009}{{\em Phys. Rev. D}
  {\bfseries 99} no.~9, (2019) 093009},
  \href{http://arxiv.org/abs/1812.05550}{{\ttfamily arXiv:1812.05550
  [hep-ph]}}.

\bibitem{Essig:2018tss}
R.~Essig, M.~Sholapurkar, and T.-T. Yu, ``{Solar Neutrinos as a Signal and
  Background in Direct-Detection Experiments Searching for Sub-GeV Dark Matter
  With Electron Recoils},''
  \href{http://dx.doi.org/10.1103/PhysRevD.97.095029}{{\em Phys. Rev. D}
  {\bfseries 97} no.~9, (2018) 095029},
  \href{http://arxiv.org/abs/1801.10159}{{\ttfamily arXiv:1801.10159
  [hep-ph]}}.

\bibitem{Boehm:2020ltd}
C.~Boehm, D.~G. Cerdeno, M.~Fairbairn, P.~A.~N. Machado, and A.~C. Vincent,
  ``{Light new physics in XENON1T},''
  \href{http://dx.doi.org/10.1103/PhysRevD.102.115013}{{\em Phys. Rev. D}
  {\bfseries 102} (2020) 115013},
  \href{http://arxiv.org/abs/2006.11250}{{\ttfamily arXiv:2006.11250
  [hep-ph]}}.

\bibitem{AristizabalSierra:2020edu}
D.~Aristizabal~Sierra, V.~De~Romeri, L.~J. Flores, and D.~K. Papoulias,
  ``{Light vector mediators facing XENON1T data},''
  \href{http://dx.doi.org/10.1016/j.physletb.2020.135681}{{\em Phys. Lett. B}
  {\bfseries 809} (2020) 135681},
  \href{http://arxiv.org/abs/2006.12457}{{\ttfamily arXiv:2006.12457
  [hep-ph]}}.

\bibitem{AristizabalSierra:2020zod}
D.~Aristizabal~Sierra, R.~Branada, O.~G. Miranda, and G.~Sanchez~Garcia,
  ``{Sensitivity of direct detection experiments to neutrino magnetic dipole
  moments},'' \href{http://dx.doi.org/10.1007/JHEP12(2020)178}{{\em JHEP}
  {\bfseries 12} (2020) 178}, \href{http://arxiv.org/abs/2008.05080}{{\ttfamily
  arXiv:2008.05080 [hep-ph]}}.

\bibitem{Amaral:2020tga}
D.~W. P.~d. Amaral, D.~G. Cerdeno, P.~Foldenauer, and E.~Reid, ``{Solar
  neutrino probes of the muon anomalous magnetic moment in the gauged $
  \mathrm{U}{(1)}_{L_{\mu }-{L}_{\tau }} $},''
  \href{http://dx.doi.org/10.1007/JHEP12(2020)155}{{\em JHEP} {\bfseries 12}
  (2020) 155}, \href{http://arxiv.org/abs/2006.11225}{{\ttfamily
  arXiv:2006.11225 [hep-ph]}}.

\bibitem{Dutta:2020che}
B.~Dutta, R.~F. Lang, S.~Liao, S.~Sinha, L.~Strigari, and A.~Thompson, ``{A
  global analysis strategy to resolve neutrino NSI degeneracies with scattering
  and oscillation data},''
  \href{http://dx.doi.org/10.1007/JHEP09(2020)106}{{\em JHEP} {\bfseries 09}
  (2020) 106}, \href{http://arxiv.org/abs/2002.03066}{{\ttfamily
  arXiv:2002.03066 [hep-ph]}}.

\bibitem{Amaral:2021rzw}
D.~W.~P. Amaral, D.~G. Cerdeno, A.~Cheek, and P.~Foldenauer, ``{Confirming
  $U(1)_{L_\mu -L_{\tau }}$ as a solution for $(g-2)_\mu $ with neutrinos},''
  \href{http://dx.doi.org/10.1140/epjc/s10052-021-09670-z}{{\em Eur. Phys. J.
  C} {\bfseries 81} no.~10, (2021) 861},
  \href{http://arxiv.org/abs/2104.03297}{{\ttfamily arXiv:2104.03297
  [hep-ph]}}.

\bibitem{DeRomeri:2024dbv}
V.~De~Romeri, D.~K. Papoulias, and C.~A. Ternes, ``{Light vector mediators at
  direct detection experiments},''
  \href{http://dx.doi.org/10.1007/JHEP05(2024)165}{{\em JHEP} {\bfseries 05}
  (2024) 165}, \href{http://arxiv.org/abs/2402.05506}{{\ttfamily
  arXiv:2402.05506 [hep-ph]}}.

\bibitem{Aalbers:2022dzr}
J.~Aalbers {\em et~al.}, ``{A next-generation liquid xenon observatory for dark
  matter and neutrino physics},''
  \href{http://dx.doi.org/10.1088/1361-6471/ac841a}{{\em J. Phys. G} {\bfseries
  50} no.~1, (2023) 013001}, \href{http://arxiv.org/abs/2203.02309}{{\ttfamily
  arXiv:2203.02309 [physics.ins-det]}}.

\bibitem{Alonso-Gonzalez:2023tgm}
D.~Alonso-Gonz\'alez, D.~W.~P. Amaral, A.~Bariego-Quintana, D.~Cerde\~no, and
  M.~de~los Rios, ``{Measuring the sterile neutrino mass in spallation source
  and direct detection experiments},''
  \href{http://dx.doi.org/10.1007/JHEP12(2023)096}{{\em JHEP} {\bfseries 12}
  (2023) 096}, \href{http://arxiv.org/abs/2307.05176}{{\ttfamily
  arXiv:2307.05176 [hep-ph]}}.

\bibitem{Amaral:2023tbs}
D.~W.~P. Amaral, D.~Cerdeno, A.~Cheek, and P.~Foldenauer, ``{A direct detection
  view of the neutrino NSI landscape},''
  \href{http://dx.doi.org/10.1007/JHEP07(2023)071}{{\em JHEP} {\bfseries 07}
  (2023) 071}, \href{http://arxiv.org/abs/2302.12846}{{\ttfamily
  arXiv:2302.12846 [hep-ph]}}.

\bibitem{Majumdar:2024dms}
A.~Majumdar, D.~K. Papoulias, H.~Prajapati, and R.~Srivastava, ``{Constraining
  low scale Dark Hypercharge symmetry at spallation, reactor and Dark Matter
  direct detection experiments},''
  \href{http://arxiv.org/abs/2411.04197}{{\ttfamily arXiv:2411.04197
  [hep-ph]}}.

\bibitem{XENON:2024ijk}
{\bfseries XENON} Collaboration, E.~Aprile {\em et~al.}, ``{First Indication of
  Solar $^8B$~Neutrinos via Coherent Elastic Neutrino-Nucleus Scattering with
  XENONnT},'' \href{http://dx.doi.org/10.1103/PhysRevLett.133.191002}{{\em
  Phys. Rev. Lett.} {\bfseries 133} (2024) 191002},
  \href{http://arxiv.org/abs/2408.02877}{{\ttfamily arXiv:2408.02877
  [nucl-ex]}}.

\bibitem{PandaX:2024muv}
{\bfseries PandaX} Collaboration, Z.~Bo {\em et~al.}, ``{First Indication of
  Solar B8 Neutrinos through Coherent Elastic Neutrino-Nucleus Scattering in
  PandaX-4T},'' \href{http://dx.doi.org/10.1103/PhysRevLett.133.191001}{{\em
  Phys. Rev. Lett.} {\bfseries 133} no.~19, (2024) 191001},
  \href{http://arxiv.org/abs/2407.10892}{{\ttfamily arXiv:2407.10892
  [hep-ex]}}.

\bibitem{XENON:2020gfr}
{\bfseries XENON} Collaboration, E.~Aprile {\em et~al.}, ``{Search for Coherent
  Elastic Scattering of Solar $^8$B Neutrinos in the XENON1T Dark Matter
  Experiment},'' \href{http://dx.doi.org/10.1103/PhysRevLett.126.091301}{{\em
  Phys. Rev. Lett.} {\bfseries 126} (2021) 091301},
  \href{http://arxiv.org/abs/2012.02846}{{\ttfamily arXiv:2012.02846
  [hep-ex]}}.

\bibitem{PandaX:2022aac}
{\bfseries PandaX} Collaboration, W.~Ma {\em et~al.}, ``{Search for Solar B8
  Neutrinos in the PandaX-4T Experiment Using Neutrino-Nucleus Coherent
  Scattering},'' \href{http://dx.doi.org/10.1103/PhysRevLett.130.021802}{{\em
  Phys. Rev. Lett.} {\bfseries 130} no.~2, (2023) 021802},
  \href{http://arxiv.org/abs/2207.04883}{{\ttfamily arXiv:2207.04883
  [hep-ex]}}.

\bibitem{Vinyoles:2016djt}
N.~Vinyoles, A.~M. Serenelli, F.~L. Villante, S.~Basu, J.~Bergstr\"om, M.~C.
  Gonzalez-Garcia, M.~Maltoni, C.~Pe\~na Garay, and N.~Song, ``{A new
  Generation of Standard Solar Models},''
  \href{http://dx.doi.org/10.3847/1538-4357/835/2/202}{{\em Astrophys. J.}
  {\bfseries 835} no.~2, (2017) 202},
  \href{http://arxiv.org/abs/1611.09867}{{\ttfamily arXiv:1611.09867
  [astro-ph.SR]}}.

\bibitem{SNO:2011hxd}
{\bfseries SNO} Collaboration, B.~Aharmim {\em et~al.}, ``{Combined Analysis of
  all Three Phases of Solar Neutrino Data from the Sudbury Neutrino
  Observatory},'' \href{http://dx.doi.org/10.1103/PhysRevC.88.025501}{{\em
  Phys. Rev. C} {\bfseries 88} (2013) 025501},
  \href{http://arxiv.org/abs/1109.0763}{{\ttfamily arXiv:1109.0763 [nucl-ex]}}.

\bibitem{KamLAND:2011fld}
{\bfseries KamLAND} Collaboration, S.~Abe {\em et~al.}, ``{Measurement of the
  8B Solar Neutrino Flux with the KamLAND Liquid Scintillator Detector},''
  \href{http://dx.doi.org/10.1103/PhysRevC.84.035804}{{\em Phys. Rev. C}
  {\bfseries 84} (2011) 035804},
  \href{http://arxiv.org/abs/1106.0861}{{\ttfamily arXiv:1106.0861 [hep-ex]}}.

\bibitem{Borexino:2017uhp}
{\bfseries Borexino} Collaboration, M.~Agostini {\em et~al.}, ``{Improved
  measurement of $^8$B solar neutrinos with $1.5 kt·y$ of Borexino
  exposure},'' \href{http://dx.doi.org/10.1103/PhysRevD.101.062001}{{\em Phys.
  Rev. D} {\bfseries 101} no.~6, (2020) 062001},
  \href{http://arxiv.org/abs/1709.00756}{{\ttfamily arXiv:1709.00756
  [hep-ex]}}.

\bibitem{Super-Kamiokande:2016yck}
{\bfseries Super-Kamiokande} Collaboration, K.~Abe {\em et~al.}, ``{Solar
  Neutrino Measurements in Super-Kamiokande-IV},''
  \href{http://dx.doi.org/10.1103/PhysRevD.94.052010}{{\em Phys. Rev. D}
  {\bfseries 94} no.~5, (2016) 052010},
  \href{http://arxiv.org/abs/1606.07538}{{\ttfamily arXiv:1606.07538
  [hep-ex]}}.

\bibitem{AristizabalSierra:2024nwf}
D.~Aristizabal~Sierra, N.~Mishra, and L.~Strigari, ``{Implications of first
  neutrino-induced nuclear recoil measurements in direct detection
  experiments},'' \href{http://arxiv.org/abs/2409.02003}{{\ttfamily
  arXiv:2409.02003 [hep-ph]}}.

\bibitem{Li:2024iij}
G.~Li, C.-Q. Song, F.-J. Tang, and J.-H. Yu, ``{Constraints on neutrino
  non-standard interactions from COHERENT and PandaX-4T},''
  \href{http://arxiv.org/abs/2409.04703}{{\ttfamily arXiv:2409.04703
  [hep-ph]}}.

\bibitem{Xia:2024ytb}
S.-y. Xia, ``{Measuring Solar neutrino Fluxes in Direct Detection Experiments
  in the Presence of Light Mediators},''
  \href{http://arxiv.org/abs/2410.01167}{{\ttfamily arXiv:2410.01167
  [hep-ph]}}.

\bibitem{Maity:2024aji}
T.~N. Maity and C.~Boehm, ``{First measurement of the weak mixing angle in
  direct detection experiments},''
  \href{http://arxiv.org/abs/2409.04385}{{\ttfamily arXiv:2409.04385
  [hep-ph]}}.

\bibitem{Lee:1956qn}
T.~D. Lee and C.-N. Yang, ``{Question of Parity Conservation in Weak
  Interactions},'' \href{http://dx.doi.org/10.1103/PhysRev.104.254}{{\em Phys.
  Rev.} {\bfseries 104} (1956) 254--258}.

\bibitem{Rodejohann:2017vup}
W.~Rodejohann, X.-J. Xu, and C.~E. Yaguna, ``{Distinguishing between Dirac and
  Majorana neutrinos in the presence of general interactions},''
  \href{http://dx.doi.org/10.1007/JHEP05(2017)024}{{\em JHEP} {\bfseries 05}
  (2017) 024}, \href{http://arxiv.org/abs/1702.05721}{{\ttfamily
  arXiv:1702.05721 [hep-ph]}}.

\bibitem{Lindner:2016wff}
M.~Lindner, W.~Rodejohann, and X.-J. Xu, ``{Coherent Neutrino-Nucleus
  Scattering and new Neutrino Interactions},''
  \href{http://dx.doi.org/10.1007/JHEP03(2017)097}{{\em JHEP} {\bfseries 03}
  (2017) 097}, \href{http://arxiv.org/abs/1612.04150}{{\ttfamily
  arXiv:1612.04150 [hep-ph]}}.

\bibitem{AristizabalSierra:2018eqm}
D.~Aristizabal~Sierra, V.~De~Romeri, and N.~Rojas, ``{COHERENT analysis of
  neutrino generalized interactions},''
  \href{http://dx.doi.org/10.1103/PhysRevD.98.075018}{{\em Phys. Rev. D}
  {\bfseries 98} (2018) 075018},
  \href{http://arxiv.org/abs/1806.07424}{{\ttfamily arXiv:1806.07424
  [hep-ph]}}.

\bibitem{Chen:2021uuw}
Z.~Chen, T.~Li, and J.~Liao, ``{Constraints on general neutrino interactions
  with exotic fermion from neutrino-electron scattering experiments},''
  \href{http://dx.doi.org/10.1007/JHEP05(2021)131}{{\em JHEP} {\bfseries 05}
  (2021) 131}, \href{http://arxiv.org/abs/2102.09784}{{\ttfamily
  arXiv:2102.09784 [hep-ph]}}.

\bibitem{Bertuzzo:2017tuf}
E.~Bertuzzo, F.~F. Deppisch, S.~Kulkarni, Y.~F. Perez~Gonzalez, and
  R.~Zukanovich~Funchal, ``{Dark Matter and Exotic Neutrino Interactions in
  Direct Detection Searches},''
  \href{http://dx.doi.org/10.1007/JHEP04(2017)073}{{\em JHEP} {\bfseries 04}
  (2017) 073}, \href{http://arxiv.org/abs/1701.07443}{{\ttfamily
  arXiv:1701.07443 [hep-ph]}}.

\bibitem{Farzan:2018gtr}
Y.~Farzan, M.~Lindner, W.~Rodejohann, and X.-J. Xu, ``{Probing neutrino
  coupling to a light scalar with coherent neutrino scattering},''
  \href{http://dx.doi.org/10.1007/JHEP05(2018)066}{{\em JHEP} {\bfseries 05}
  (2018) 066}, \href{http://arxiv.org/abs/1802.05171}{{\ttfamily
  arXiv:1802.05171 [hep-ph]}}.

\bibitem{Denton:2022nol}
P.~B. Denton and J.~Gehrlein, ``{New constraints on the dark side of
  non-standard interactions from reactor neutrino scattering data},''
  \href{http://dx.doi.org/10.1103/PhysRevD.106.015022}{{\em Phys. Rev. D}
  {\bfseries 106} no.~1, (2022) 015022},
  \href{http://arxiv.org/abs/2204.09060}{{\ttfamily arXiv:2204.09060
  [hep-ph]}}.

\bibitem{Barranco:2005yy}
J.~Barranco, O.~G. Miranda, and T.~I. Rashba, ``{Probing new physics with
  coherent neutrino scattering off nuclei},''
  \href{http://dx.doi.org/10.1088/1126-6708/2005/12/021}{{\em JHEP} {\bfseries
  12} (2005) 021}, \href{http://arxiv.org/abs/hep-ph/0508299}{{\ttfamily
  arXiv:hep-ph/0508299}}.

\bibitem{Cadeddu:2020lky}
M.~Cadeddu, F.~Dordei, C.~Giunti, Y.~F. Li, E.~Picciau, and Y.~Y. Zhang,
  ``{Physics results from the first COHERENT observation of coherent elastic
  neutrino-nucleus scattering in argon and their combination with cesium-iodide
  data},'' \href{http://dx.doi.org/10.1103/PhysRevD.102.015030}{{\em Phys. Rev.
  D} {\bfseries 102} no.~1, (2020) 015030},
  \href{http://arxiv.org/abs/2005.01645}{{\ttfamily arXiv:2005.01645
  [hep-ph]}}.

\bibitem{ParticleDataGroup:2024cfk}
{\bfseries Particle Data Group} Collaboration, S.~Navas {\em et~al.}, ``{Review
  of particle physics},''
  \href{http://dx.doi.org/10.1103/PhysRevD.110.030001}{{\em Phys. Rev. D}
  {\bfseries 110} no.~3, (2024) 030001}.

\bibitem{Klein:1999qj}
S.~Klein and J.~Nystrand, ``{Exclusive vector meson production in relativistic
  heavy ion collisions},''
  \href{http://dx.doi.org/10.1103/PhysRevC.60.014903}{{\em Phys. Rev. C}
  {\bfseries 60} (1999) 014903},
  \href{http://arxiv.org/abs/hep-ph/9902259}{{\ttfamily arXiv:hep-ph/9902259}}.

\bibitem{Candela:2024ljb}
P.~M. Candela, V.~De~Romeri, P.~Melas, D.~K. Papoulias, and N.~Saoulidou,
  ``{Up-scattering production of a sterile fermion at DUNE: complementarity
  with spallation source and direct detection experiments},''
  \href{http://dx.doi.org/10.1007/JHEP10(2024)032}{{\em JHEP} {\bfseries 10}
  (2024) 032}, \href{http://arxiv.org/abs/2404.12476}{{\ttfamily
  arXiv:2404.12476 [hep-ph]}}.

\bibitem{Langacker:2008yv}
P.~Langacker, ``{The Physics of Heavy $Z^\prime$ Gauge Bosons},''
  \href{http://dx.doi.org/10.1103/RevModPhys.81.1199}{{\em Rev. Mod. Phys.}
  {\bfseries 81} (2009) 1199--1228},
  \href{http://arxiv.org/abs/0801.1345}{{\ttfamily arXiv:0801.1345 [hep-ph]}}.

\bibitem{Okada:2018ktp}
S.~Okada, ``{$Z'$ Portal Dark Matter in the Minimal $B-L$ Model},''
  \href{http://dx.doi.org/10.1155/2018/5340935}{{\em Adv. High Energy Phys.}
  {\bfseries 2018} (2018) 5340935},
  \href{http://arxiv.org/abs/1803.06793}{{\ttfamily arXiv:1803.06793
  [hep-ph]}}.

\bibitem{Cirelli:2013ufw}
M.~Cirelli, E.~Del~Nobile, and P.~Panci, ``{Tools for model-independent bounds
  in direct dark matter searches},''
  \href{http://dx.doi.org/10.1088/1475-7516/2013/10/019}{{\em JCAP} {\bfseries
  10} (2013) 019}, \href{http://arxiv.org/abs/1307.5955}{{\ttfamily
  arXiv:1307.5955 [hep-ph]}}.

\bibitem{DelNobile:2021wmp}
E.~Del~Nobile, ``{The Theory of Direct Dark Matter Detection: A Guide to
  Computations},'' \href{http://arxiv.org/abs/2104.12785}{{\ttfamily
  arXiv:2104.12785 [hep-ph]}}.

\bibitem{Hoferichter:2020osn}
M.~Hoferichter, J.~Men\'endez, and A.~Schwenk, ``{Coherent elastic
  neutrino-nucleus scattering: EFT analysis and nuclear responses},''
  \href{http://dx.doi.org/10.1103/PhysRevD.102.074018}{{\em Phys. Rev. D}
  {\bfseries 102} no.~7, (2020) 074018},
  \href{http://arxiv.org/abs/2007.08529}{{\ttfamily arXiv:2007.08529
  [hep-ph]}}.

\bibitem{XENON:2024hup}
{\bfseries XENON} Collaboration, E.~Aprile {\em et~al.}, ``{First Search for
  Light Dark Matter in the Neutrino Fog with XENONnT},''
  \href{http://arxiv.org/abs/2409.17868}{{\ttfamily arXiv:2409.17868
  [hep-ex]}}.

\bibitem{bahcall_web}
J.~Bahcall. \url{http://www.sns.ias.edu/~jnb/SNdata/sndata.html}.

\bibitem{Bahcall:1996qv}
J.~N. Bahcall, E.~Lisi, D.~E. Alburger, L.~De~Braeckeleer, S.~J. Freedman, and
  J.~Napolitano, ``{Standard neutrino spectrum from B-8 decay},''
  \href{http://dx.doi.org/10.1103/PhysRevC.54.411}{{\em Phys. Rev. C}
  {\bfseries 54} (1996) 411--422},
  \href{http://arxiv.org/abs/nucl-th/9601044}{{\ttfamily
  arXiv:nucl-th/9601044}}.

\bibitem{XENON:2024xgd}
{\bfseries XENON} Collaboration, E.~Aprile {\em et~al.}, ``{XENONnT WIMP
  Search: Signal \& Background Modeling and Statistical Inference},''
  \href{http://arxiv.org/abs/2406.13638}{{\ttfamily arXiv:2406.13638
  [physics.data-an]}}.

\bibitem{Erler:2004in}
J.~Erler and M.~J. Ramsey-Musolf, ``{The Weak mixing angle at low energies},''
  \href{http://dx.doi.org/10.1103/PhysRevD.72.073003}{{\em Phys. Rev. D}
  {\bfseries 72} (2005) 073003},
  \href{http://arxiv.org/abs/hep-ph/0409169}{{\ttfamily arXiv:hep-ph/0409169}}.

\bibitem{Wood:1997zq}
C.~S. Wood, S.~C. Bennett, D.~Cho, B.~P. Masterson, J.~L. Roberts, C.~E.
  Tanner, and C.~E. Wieman, ``{Measurement of parity nonconservation and an
  anapole moment in cesium},''
  \href{http://dx.doi.org/10.1126/science.275.5307.1759}{{\em Science}
  {\bfseries 275} (1997) 1759--1763}.

\bibitem{Qweak:2018tjf}
{\bfseries Qweak} Collaboration, D.~Androi\'c {\em et~al.}, ``{Precision
  measurement of the weak charge of the proton},''
  \href{http://dx.doi.org/10.1038/s41586-018-0096-0}{{\em Nature} {\bfseries
  557} no.~7704, (2018) 207--211},
  \href{http://arxiv.org/abs/1905.08283}{{\ttfamily arXiv:1905.08283
  [nucl-ex]}}.

\bibitem{SLACE158:2005uay}
{\bfseries SLAC E158} Collaboration, P.~L. Anthony {\em et~al.}, ``{Precision
  measurement of the weak mixing angle in Moller scattering},''
  \href{http://dx.doi.org/10.1103/PhysRevLett.95.081601}{{\em Phys. Rev. Lett.}
  {\bfseries 95} (2005) 081601},
  \href{http://arxiv.org/abs/hep-ex/0504049}{{\ttfamily arXiv:hep-ex/0504049}}.

\bibitem{PVDIS:2014cmd}
{\bfseries PVDIS} Collaboration, D.~Wang {\em et~al.}, ``{Measurement of parity
  violation in electron\textendash{}quark scattering},''
  \href{http://dx.doi.org/10.1038/nature12964}{{\em Nature} {\bfseries 506}
  no.~7486, (2014) 67--70}.

\bibitem{NuTeV:2001whx}
{\bfseries NuTeV} Collaboration, G.~P. Zeller {\em et~al.}, ``{A Precise
  Determination of Electroweak Parameters in Neutrino Nucleon Scattering},''
  \href{http://dx.doi.org/10.1103/PhysRevLett.88.091802}{{\em Phys. Rev. Lett.}
  {\bfseries 88} (2002) 091802},
  \href{http://arxiv.org/abs/hep-ex/0110059}{{\ttfamily arXiv:hep-ex/0110059}}.
  [Erratum: Phys.Rev.Lett. 90, 239902 (2003)].

\bibitem{DeRomeri:2022twg}
V.~De~Romeri, O.~G. Miranda, D.~K. Papoulias, G.~Sanchez~Garcia, M.~T\'ortola,
  and J.~W.~F. Valle, ``{Physics implications of a combined analysis of
  COHERENT CsI and LAr data},''
  \href{http://dx.doi.org/10.1007/JHEP04(2023)035}{{\em JHEP} {\bfseries 04}
  (2023) 035}, \href{http://arxiv.org/abs/2211.11905}{{\ttfamily
  arXiv:2211.11905 [hep-ph]}}.

\bibitem{AristizabalSierra:2022axl}
D.~Aristizabal~Sierra, V.~De~Romeri, and D.~K. Papoulias, ``{Consequences of
  the Dresden-II reactor data for the weak mixing angle and new physics},''
  \href{http://dx.doi.org/10.1007/JHEP09(2022)076}{{\em JHEP} {\bfseries 09}
  (2022) 076}, \href{http://arxiv.org/abs/2203.02414}{{\ttfamily
  arXiv:2203.02414 [hep-ph]}}.

\bibitem{Majumdar:2022nby}
A.~Majumdar, D.~K. Papoulias, R.~Srivastava, and J.~W.~F. Valle, ``{Physics
  implications of recent Dresden-II reactor data},''
  \href{http://dx.doi.org/10.1103/PhysRevD.106.093010}{{\em Phys. Rev. D}
  {\bfseries 106} no.~9, (2022) 093010},
  \href{http://arxiv.org/abs/2208.13262}{{\ttfamily arXiv:2208.13262
  [hep-ph]}}.

\bibitem{AtzoriCorona:2024vhj}
M.~Atzori~Corona, M.~Cadeddu, N.~Cargioli, F.~Dordei, and C.~Giunti, ``{Refined
  determination of the weak mixing angle at low energy},''
  \href{http://dx.doi.org/10.1103/PhysRevD.110.033005}{{\em Phys. Rev. D}
  {\bfseries 110} no.~3, (2024) 033005},
  \href{http://arxiv.org/abs/2405.09416}{{\ttfamily arXiv:2405.09416
  [hep-ph]}}.

\bibitem{AtzoriCorona:2022qrf}
M.~Atzori~Corona, M.~Cadeddu, N.~Cargioli, F.~Dordei, C.~Giunti, Y.~F. Li,
  C.~A. Ternes, and Y.~Y. Zhang, ``{Impact of the Dresden-II and COHERENT
  neutrino scattering data on neutrino electromagnetic properties and
  electroweak physics},'' \href{http://dx.doi.org/10.1007/JHEP09(2022)164}{{\em
  JHEP} {\bfseries 09} (2022) 164},
  \href{http://arxiv.org/abs/2205.09484}{{\ttfamily arXiv:2205.09484
  [hep-ph]}}.

\bibitem{AtzoriCorona:2023ktl}
M.~Atzori~Corona, M.~Cadeddu, N.~Cargioli, F.~Dordei, C.~Giunti, and G.~Masia,
  ``{Nuclear neutron radius and weak mixing angle measurements from latest
  COHERENT CsI and atomic parity violation Cs data},''
  \href{http://dx.doi.org/10.1140/epjc/s10052-023-11849-5}{{\em Eur. Phys. J.
  C} {\bfseries 83} no.~7, (2023) 683},
  \href{http://arxiv.org/abs/2303.09360}{{\ttfamily arXiv:2303.09360
  [nucl-ex]}}.

\bibitem{Cadeddu:2021ijh}
M.~Cadeddu, N.~Cargioli, F.~Dordei, C.~Giunti, Y.~F. Li, E.~Picciau, C.~A.
  Ternes, and Y.~Y. Zhang, ``{New insights into nuclear physics and weak mixing
  angle using electroweak probes},''
  \href{http://dx.doi.org/10.1103/PhysRevC.104.065502}{{\em Phys. Rev. C}
  {\bfseries 104} no.~6, (2021) 065502},
  \href{http://arxiv.org/abs/2102.06153}{{\ttfamily arXiv:2102.06153
  [hep-ph]}}.

\bibitem{Cadeddu:2024baq}
M.~Cadeddu, N.~Cargioli, J.~Erler, M.~Gorchtein, J.~Piekarewicz, X.~Roca-Maza,
  and H.~Spiesberger, ``{Simultaneous extraction of the weak radius and the
  weak mixing angle from parity-violating electron scattering on C12},''
  \href{http://dx.doi.org/10.1103/PhysRevC.110.035501}{{\em Phys. Rev. C}
  {\bfseries 110} no.~3, (2024) 035501},
  \href{http://arxiv.org/abs/2407.09743}{{\ttfamily arXiv:2407.09743
  [hep-ph]}}.

\bibitem{Alonso:2021kyu}
J.~Alonso {\em et~al.}, ``{Neutrino physics opportunities with the IsoDAR
  source at Yemilab},''
  \href{http://dx.doi.org/10.1103/PhysRevD.105.052009}{{\em Phys. Rev. D}
  {\bfseries 105} no.~5, (2022) 052009},
  \href{http://arxiv.org/abs/2111.09480}{{\ttfamily arXiv:2111.09480
  [hep-ex]}}.

\bibitem{Alves:2024twb}
G.~F.~S. Alves, A.~P. Ferreira, S.~W. Li, P.~A.~N. Machado, and Y.~F.
  Perez-Gonzalez, ``{Measuring the weak mixing angle at SBND},''
  \href{http://arxiv.org/abs/2409.07430}{{\ttfamily arXiv:2409.07430
  [hep-ph]}}.

\bibitem{deGouvea:2019wav}
A.~de~Gouvea, P.~A.~N. Machado, Y.~F. Perez-Gonzalez, and Z.~Tabrizi,
  ``{Measuring the Weak Mixing Angle in the DUNE Near Detector Complex},''
  \href{http://dx.doi.org/10.1103/PhysRevLett.125.051803}{{\em Phys. Rev.
  Lett.} {\bfseries 125} no.~5, (2020) 051803},
  \href{http://arxiv.org/abs/1912.06658}{{\ttfamily arXiv:1912.06658
  [hep-ph]}}.

\bibitem{AtzoriCorona:2022moj}
M.~Atzori~Corona, M.~Cadeddu, N.~Cargioli, F.~Dordei, C.~Giunti, Y.~F. Li,
  E.~Picciau, C.~A. Ternes, and Y.~Y. Zhang, ``{Probing light mediators and (g
  \ensuremath{-} 2)$_{\mu}$ through detection of coherent elastic neutrino
  nucleus scattering at COHERENT},''
  \href{http://dx.doi.org/10.1007/JHEP05(2022)109}{{\em JHEP} {\bfseries 05}
  (2022) 109}, \href{http://arxiv.org/abs/2202.11002}{{\ttfamily
  arXiv:2202.11002 [hep-ph]}}.

\bibitem{CONUS:2021dwh}
{\bfseries CONUS} Collaboration, H.~Bonet {\em et~al.}, ``{Novel constraints on
  neutrino physics beyond the standard model from the CONUS experiment},''
  \href{http://dx.doi.org/10.1007/JHEP05(2022)085}{{\em JHEP} {\bfseries 05}
  (2022) 085}, \href{http://arxiv.org/abs/2110.02174}{{\ttfamily
  arXiv:2110.02174 [hep-ph]}}.

\bibitem{Lindner:2024eng}
M.~Lindner, T.~Rink, and M.~Sen, ``{Light vector bosons and the weak mixing
  angle in the light of future germanium-based reactor CE\ensuremath{\nu}NS
  experiments},'' \href{http://dx.doi.org/10.1007/JHEP08(2024)171}{{\em JHEP}
  {\bfseries 08} (2024) 171}, \href{http://arxiv.org/abs/2401.13025}{{\ttfamily
  arXiv:2401.13025 [hep-ph]}}.

\bibitem{CONNIE:2019xid}
{\bfseries CONNIE} Collaboration, A.~Aguilar-Arevalo {\em et~al.}, ``{Search
  for light mediators in the low-energy data of the CONNIE reactor neutrino
  experiment},'' \href{http://dx.doi.org/10.1007/JHEP04(2020)054}{{\em JHEP}
  {\bfseries 04} (2020) 054}, \href{http://arxiv.org/abs/1910.04951}{{\ttfamily
  arXiv:1910.04951 [hep-ex]}}.

\bibitem{CONNIE:2024pwt}
{\bfseries CONNIE} Collaboration, A.~A. Aguilar-Arevalo {\em et~al.},
  ``{Searches for CE\ensuremath{\nu}NS and Physics beyond the Standard Model
  using Skipper-CCDs at CONNIE},''
  \href{http://arxiv.org/abs/2403.15976}{{\ttfamily arXiv:2403.15976
  [hep-ex]}}.

\bibitem{Coloma:2022avw}
P.~Coloma, I.~Esteban, M.~C. Gonzalez-Garcia, L.~Larizgoitia, F.~Monrabal, and
  S.~Palomares-Ruiz, ``{Bounds on new physics with data of the Dresden-II
  reactor experiment and COHERENT},''
  \href{http://dx.doi.org/10.1007/JHEP05(2022)037}{{\em JHEP} {\bfseries 05}
  (2022) 037}, \href{http://arxiv.org/abs/2202.10829}{{\ttfamily
  arXiv:2202.10829 [hep-ph]}}.

\bibitem{Bauer:2018onh}
M.~Bauer, P.~Foldenauer, and J.~Jaeckel, ``{Hunting All the Hidden Photons},''
  \href{http://dx.doi.org/10.1007/JHEP07(2018)094}{{\em JHEP} {\bfseries 07}
  (2018) 094}, \href{http://arxiv.org/abs/1803.05466}{{\ttfamily
  arXiv:1803.05466 [hep-ph]}}.

\bibitem{TEXONO:2009knm}
{\bfseries TEXONO} Collaboration, M.~Deniz {\em et~al.}, ``{Measurement of
  Nu(e)-bar -Electron Scattering Cross-Section with a CsI(Tl) Scintillating
  Crystal Array at the Kuo-Sheng Nuclear Power Reactor},''
  \href{http://dx.doi.org/10.1103/PhysRevD.81.072001}{{\em Phys. Rev. D}
  {\bfseries 81} (2010) 072001},
  \href{http://arxiv.org/abs/0911.1597}{{\ttfamily arXiv:0911.1597 [hep-ex]}}.

\bibitem{A:2022acy}
S.~K. A., A.~Majumdar, D.~K. Papoulias, H.~Prajapati, and R.~Srivastava,
  ``{Implications of first LZ and XENONnT results: A comparative study of
  neutrino properties and light mediators},''
  \href{http://dx.doi.org/10.1016/j.physletb.2023.137742}{{\em Phys. Lett. B}
  {\bfseries 839} (2023) 137742},
  \href{http://arxiv.org/abs/2208.06415}{{\ttfamily arXiv:2208.06415
  [hep-ph]}}.

\bibitem{Riordan:1987aw}
E.~M. Riordan {\em et~al.}, ``{A Search for Short Lived Axions in an Electron
  Beam Dump Experiment},''
  \href{http://dx.doi.org/10.1103/PhysRevLett.59.755}{{\em Phys. Rev. Lett.}
  {\bfseries 59} (1987) 755}.

\bibitem{Bjorken:1988as}
J.~D. Bjorken, S.~Ecklund, W.~R. Nelson, A.~Abashian, C.~Church, B.~Lu, L.~W.
  Mo, T.~A. Nunamaker, and P.~Rassmann, ``{Search for Neutral Metastable
  Penetrating Particles Produced in the SLAC Beam Dump},''
  \href{http://dx.doi.org/10.1103/PhysRevD.38.3375}{{\em Phys. Rev. D}
  {\bfseries 38} (1988) 3375}.

\bibitem{Bross:1989mp}
A.~Bross, M.~Crisler, S.~H. Pordes, J.~Volk, S.~Errede, and J.~Wrbanek, ``{A
  Search for Shortlived Particles Produced in an Electron Beam Dump},''
  \href{http://dx.doi.org/10.1103/PhysRevLett.67.2942}{{\em Phys. Rev. Lett.}
  {\bfseries 67} (1991) 2942--2945}.

\bibitem{Konaka:1986cb}
A.~Konaka {\em et~al.}, ``{Search for Neutral Particles in Electron Beam Dump
  Experiment},'' \href{http://dx.doi.org/10.1103/PhysRevLett.57.659}{{\em Phys.
  Rev. Lett.} {\bfseries 57} (1986) 659}.

\bibitem{Davier:1989wz}
M.~Davier and H.~Nguyen~Ngoc, ``{An Unambiguous Search for a Light Higgs
  Boson},'' \href{http://dx.doi.org/10.1016/0370-2693(89)90174-3}{{\em Phys.
  Lett. B} {\bfseries 229} (1989) 150--155}.

\bibitem{Bjorken:2009mm}
J.~D. Bjorken, R.~Essig, P.~Schuster, and N.~Toro, ``{New Fixed-Target
  Experiments to Search for Dark Gauge Forces},''
  \href{http://dx.doi.org/10.1103/PhysRevD.80.075018}{{\em Phys. Rev. D}
  {\bfseries 80} (2009) 075018},
  \href{http://arxiv.org/abs/0906.0580}{{\ttfamily arXiv:0906.0580 [hep-ph]}}.

\bibitem{Andreas:2012mt}
S.~Andreas, C.~Niebuhr, and A.~Ringwald, ``{New Limits on Hidden Photons from
  Past Electron Beam Dumps},''
  \href{http://dx.doi.org/10.1103/PhysRevD.86.095019}{{\em Phys. Rev. D}
  {\bfseries 86} (2012) 095019},
  \href{http://arxiv.org/abs/1209.6083}{{\ttfamily arXiv:1209.6083 [hep-ph]}}.

\bibitem{Blumlein:1990ay}
J.~Blumlein {\em et~al.}, ``{Limits on neutral light scalar and pseudoscalar
  particles in a proton beam dump experiment},''
  \href{http://dx.doi.org/10.1007/BF01548556}{{\em Z. Phys. C} {\bfseries 51}
  (1991) 341--350}.

\bibitem{Blumlein:1991xh}
J.~Blumlein {\em et~al.}, ``{Limits on the mass of light (pseudo)scalar
  particles from Bethe-Heitler e+ e- and mu+ mu- pair production in a proton -
  iron beam dump experiment},''
  \href{http://dx.doi.org/10.1142/S0217751X9200171X}{{\em Int. J. Mod. Phys. A}
  {\bfseries 7} (1992) 3835--3850}.

\bibitem{Blumlein:2011mv}
J.~Blumlein and J.~Brunner, ``{New Exclusion Limits for Dark Gauge Forces from
  Beam-Dump Data},''
  \href{http://dx.doi.org/10.1016/j.physletb.2011.05.046}{{\em Phys. Lett. B}
  {\bfseries 701} (2011) 155--159},
  \href{http://arxiv.org/abs/1104.2747}{{\ttfamily arXiv:1104.2747 [hep-ex]}}.

\bibitem{Blumlein:2013cua}
J.~Bl\"umlein and J.~Brunner, ``{New Exclusion Limits on Dark Gauge Forces from
  Proton Bremsstrahlung in Beam-Dump Data},''
  \href{http://dx.doi.org/10.1016/j.physletb.2014.02.029}{{\em Phys. Lett. B}
  {\bfseries 731} (2014) 320--326},
  \href{http://arxiv.org/abs/1311.3870}{{\ttfamily arXiv:1311.3870 [hep-ph]}}.

\bibitem{CHARM:1985anb}
{\bfseries CHARM} Collaboration, F.~Bergsma {\em et~al.}, ``{Search for Axion
  Like Particle Production in 400-{GeV} Proton - Copper Interactions},''
  \href{http://dx.doi.org/10.1016/0370-2693(85)90400-9}{{\em Phys. Lett. B}
  {\bfseries 157} (1985) 458--462}.

\bibitem{Gninenko:2012eq}
S.~N. Gninenko, ``{Constraints on sub-GeV hidden sector gauge bosons from a
  search for heavy neutrino decays},''
  \href{http://dx.doi.org/10.1016/j.physletb.2012.06.002}{{\em Phys. Lett. B}
  {\bfseries 713} (2012) 244--248},
  \href{http://arxiv.org/abs/1204.3583}{{\ttfamily arXiv:1204.3583 [hep-ph]}}.

\bibitem{NOMAD:2001eyx}
{\bfseries NOMAD} Collaboration, P.~Astier {\em et~al.}, ``{Search for heavy
  neutrinos mixing with tau neutrinos},''
  \href{http://dx.doi.org/10.1016/S0370-2693(01)00362-8}{{\em Phys. Lett. B}
  {\bfseries 506} (2001) 27--38},
  \href{http://arxiv.org/abs/hep-ex/0101041}{{\ttfamily arXiv:hep-ex/0101041}}.

\bibitem{Bernardi:1985ny}
G.~Bernardi {\em et~al.}, ``{Search for Neutrino Decay},''
  \href{http://dx.doi.org/10.1016/0370-2693(86)91602-3}{{\em Phys. Lett. B}
  {\bfseries 166} (1986) 479--483}.

\bibitem{Gninenko:2011uv}
S.~N. Gninenko, ``{Stringent limits on the $\pi^0 \to \gamma X, X \to e+e-$
  decay from neutrino experiments and constraints on new light gauge bosons},''
  \href{http://dx.doi.org/10.1103/PhysRevD.85.055027}{{\em Phys. Rev. D}
  {\bfseries 85} (2012) 055027},
  \href{http://arxiv.org/abs/1112.5438}{{\ttfamily arXiv:1112.5438 [hep-ph]}}.

\bibitem{Merkel:2014avp}
H.~Merkel {\em et~al.}, ``{Search at the Mainz Microtron for Light Massive
  Gauge Bosons Relevant for the Muon g-2 Anomaly},''
  \href{http://dx.doi.org/10.1103/PhysRevLett.112.221802}{{\em Phys. Rev.
  Lett.} {\bfseries 112} no.~22, (2014) 221802},
  \href{http://arxiv.org/abs/1404.5502}{{\ttfamily arXiv:1404.5502 [hep-ex]}}.

\bibitem{APEX:2011dww}
{\bfseries APEX} Collaboration, S.~Abrahamyan {\em et~al.}, ``{Search for a New
  Gauge Boson in Electron-Nucleus Fixed-Target Scattering by the APEX
  Experiment},'' \href{http://dx.doi.org/10.1103/PhysRevLett.107.191804}{{\em
  Phys. Rev. Lett.} {\bfseries 107} (2011) 191804},
  \href{http://arxiv.org/abs/1108.2750}{{\ttfamily arXiv:1108.2750 [hep-ex]}}.

\bibitem{BaBar:2014zli}
{\bfseries BaBar} Collaboration, J.~P. Lees {\em et~al.}, ``{Search for a Dark
  Photon in $e^+e^-$ Collisions at BaBar},''
  \href{http://dx.doi.org/10.1103/PhysRevLett.113.201801}{{\em Phys. Rev.
  Lett.} {\bfseries 113} no.~20, (2014) 201801},
  \href{http://arxiv.org/abs/1406.2980}{{\ttfamily arXiv:1406.2980 [hep-ex]}}.

\bibitem{BaBar:2017tiz}
{\bfseries BaBar} Collaboration, J.~P. Lees {\em et~al.}, ``{Search for
  Invisible Decays of a Dark Photon Produced in ${e}^{+}{e}^{-}$ Collisions at
  BaBar},'' \href{http://dx.doi.org/10.1103/PhysRevLett.119.131804}{{\em Phys.
  Rev. Lett.} {\bfseries 119} no.~13, (2017) 131804},
  \href{http://arxiv.org/abs/1702.03327}{{\ttfamily arXiv:1702.03327
  [hep-ex]}}.

\bibitem{LHCb:2017trq}
{\bfseries LHCb} Collaboration, R.~Aaij {\em et~al.}, ``{Search for Dark
  Photons Produced in 13 TeV $pp$ Collisions},''
  \href{http://dx.doi.org/10.1103/PhysRevLett.120.061801}{{\em Phys. Rev.
  Lett.} {\bfseries 120} no.~6, (2018) 061801},
  \href{http://arxiv.org/abs/1710.02867}{{\ttfamily arXiv:1710.02867
  [hep-ex]}}.

\bibitem{NA64:2021xzo}
{\bfseries NA64} Collaboration, Y.~M. Andreev {\em et~al.}, ``{Constraints on
  New Physics in Electron $g-2$ from a Search for Invisible Decays of a Scalar,
  Pseudoscalar, Vector, and Axial Vector},''
  \href{http://dx.doi.org/10.1103/PhysRevLett.126.211802}{{\em Phys. Rev.
  Lett.} {\bfseries 126} no.~21, (2021) 211802},
  \href{http://arxiv.org/abs/2102.01885}{{\ttfamily arXiv:2102.01885
  [hep-ex]}}.

\bibitem{NA64:2022yly}
{\bfseries NA64} Collaboration, Y.~M. Andreev {\em et~al.}, ``{Search for a New
  B-L Z' Gauge Boson with the NA64 Experiment at CERN},''
  \href{http://dx.doi.org/10.1103/PhysRevLett.129.161801}{{\em Phys. Rev.
  Lett.} {\bfseries 129} no.~16, (2022) 161801},
  \href{http://arxiv.org/abs/2207.09979}{{\ttfamily arXiv:2207.09979
  [hep-ex]}}.

\bibitem{NA64:2023wbi}
{\bfseries NA64} Collaboration, Y.~M. Andreev {\em et~al.}, ``{Search for Light
  Dark Matter with NA64 at CERN},''
  \href{http://dx.doi.org/10.1103/PhysRevLett.131.161801}{{\em Phys. Rev.
  Lett.} {\bfseries 131} no.~16, (2023) 161801},
  \href{http://arxiv.org/abs/2307.02404}{{\ttfamily arXiv:2307.02404
  [hep-ex]}}.

\bibitem{Ilten:2018crw}
P.~Ilten, Y.~Soreq, M.~Williams, and W.~Xue, ``{Serendipity in dark photon
  searches},'' \href{http://dx.doi.org/10.1007/JHEP06(2018)004}{{\em JHEP}
  {\bfseries 06} (2018) 004}, \href{http://arxiv.org/abs/1801.04847}{{\ttfamily
  arXiv:1801.04847 [hep-ph]}}.

\bibitem{Baruch:2022esd}
C.~Baruch, P.~Ilten, Y.~Soreq, and M.~Williams, ``{Axial vectors in
  DarkCast},'' \href{http://dx.doi.org/10.1007/JHEP11(2022)124}{{\em JHEP}
  {\bfseries 11} (2022) 124}, \href{http://arxiv.org/abs/2206.08563}{{\ttfamily
  arXiv:2206.08563 [hep-ph]}}.

\bibitem{Esseili:2023ldf}
H.~Esseili and G.~D. Kribs, ``{Cosmological implications of gauged U(1)$_{B-L}$
  on \ensuremath{\Delta}N $_{eff}$ in the CMB and BBN},''
  \href{http://dx.doi.org/10.1088/1475-7516/2024/05/110}{{\em JCAP} {\bfseries
  05} (2024) 110}, \href{http://arxiv.org/abs/2308.07955}{{\ttfamily
  arXiv:2308.07955 [hep-ph]}}.

\bibitem{Li:2023puz}
S.-P. Li and X.-J. Xu, ``{N$_{eff}$ constraints on light mediators coupled to
  neutrinos: the dilution-resistant effect},''
  \href{http://dx.doi.org/10.1007/JHEP10(2023)012}{{\em JHEP} {\bfseries 10}
  (2023) 012}, \href{http://arxiv.org/abs/2307.13967}{{\ttfamily
  arXiv:2307.13967 [hep-ph]}}.

\bibitem{Ghosh:2024cxi}
D.~K. Ghosh, P.~Ghosh, S.~Jeesun, and R.~Srivastava, ``{Neff at CMB challenges
  U(1)X light gauge boson scenarios},''
  \href{http://dx.doi.org/10.1103/PhysRevD.110.075032}{{\em Phys. Rev. D}
  {\bfseries 110} no.~7, (2024) 075032},
  \href{http://arxiv.org/abs/2404.10077}{{\ttfamily arXiv:2404.10077
  [hep-ph]}}.

\bibitem{PhysRevD.110.075032}
D.~K. Ghosh, P.~Ghosh, S.~Jeesun, and R.~Srivastava, ``${N}_{\mathrm{eff}}$ at
  cmb challenges $u(1{)}_{X}$ light gauge boson scenarios,''
  \href{http://dx.doi.org/10.1103/PhysRevD.110.075032}{{\em Phys. Rev. D}
  {\bfseries 110} (Oct, 2024) 075032}.
  \url{https://link.aps.org/doi/10.1103/PhysRevD.110.075032}.

\bibitem{Heurtier:2016otg}
L.~Heurtier and Y.~Zhang, ``{Supernova Constraints on Massive (Pseudo)Scalar
  Coupling to Neutrinos},''
  \href{http://dx.doi.org/10.1088/1475-7516/2017/02/042}{{\em JCAP} {\bfseries
  02} (2017) 042}, \href{http://arxiv.org/abs/1609.05882}{{\ttfamily
  arXiv:1609.05882 [hep-ph]}}.

\bibitem{Chang:2016ntp}
J.~H. Chang, R.~Essig, and S.~D. McDermott, ``{Revisiting Supernova 1987A
  Constraints on Dark Photons},''
  \href{http://dx.doi.org/10.1007/JHEP01(2017)107}{{\em JHEP} {\bfseries 01}
  (2017) 107}, \href{http://arxiv.org/abs/1611.03864}{{\ttfamily
  arXiv:1611.03864 [hep-ph]}}.

\bibitem{Croon:2020lrf}
D.~Croon, G.~Elor, R.~K. Leane, and S.~D. McDermott, ``{Supernova Muons: New
  Constraints on $Z$' Bosons, Axions and ALPs},''
  \href{http://dx.doi.org/10.1007/JHEP01(2021)107}{{\em JHEP} {\bfseries 01}
  (2021) 107}, \href{http://arxiv.org/abs/2006.13942}{{\ttfamily
  arXiv:2006.13942 [hep-ph]}}.

\bibitem{Caputo:2021rux}
A.~Caputo, G.~Raffelt, and E.~Vitagliano, ``{Muonic boson limits: Supernova
  redux},'' \href{http://dx.doi.org/10.1103/PhysRevD.105.035022}{{\em Phys.
  Rev. D} {\bfseries 105} no.~3, (2022) 035022},
  \href{http://arxiv.org/abs/2109.03244}{{\ttfamily arXiv:2109.03244
  [hep-ph]}}.

\bibitem{Caputo:2022rca}
A.~Caputo, G.~Raffelt, and E.~Vitagliano, ``{Radiative transfer in stars by
  feebly interacting bosons},''
  \href{http://dx.doi.org/10.1088/1475-7516/2022/08/045}{{\em JCAP} {\bfseries
  08} no.~08, (2022) 045}, \href{http://arxiv.org/abs/2204.11862}{{\ttfamily
  arXiv:2204.11862 [astro-ph.SR]}}.

\end{thebibliography}\endgroup

\end{document}